\documentclass[journal,pdftex,onecolumn]{IEEEtran}

\usepackage{adjustbox}
\usepackage{natbib}
\ifCLASSINFOpdf
   \usepackage[pdftex]{graphicx}
\else
   \usepackage[dvips]{graphicx}
  \fi
\usepackage{wrapfig}
\usepackage{amsmath}                            
\usepackage{amssymb}                            
\usepackage{amsthm}

\usepackage{multicol}
\usepackage[letterpaper, margin = 0.75in, top = 0.75in]{geometry}
\usepackage{rotating}
\usepackage{caption}
\usepackage{longtable}
\usepackage{array}
\newcolumntype{P}[1]{>{\centering\arraybackslash}p{#1}}
\usepackage{hhline}
\usepackage{algorithm}
\usepackage{algpseudocode}
\usepackage{xparse}
\usepackage{xcolor}

\makeatletter
\NewDocumentCommand{\LeftComment}{s m}{%
  \Statex \IfBooleanF{#1}{\hspace*{\ALG@thistlm}}#2}
\makeatother

\usepackage{setspace}
\theoremstyle{definition}

\usepackage{listings}
\usepackage{xcolor}

\usepackage{amsmath}
\usepackage[colorlinks=true,
pdfpagemode=UseNone,
urlcolor=blue,
bookmarks=true,
backref=page
]{hyperref}

\begin{document}

\title{
Opening the House: Datasets for Mixed Doubles Curling
} 

\author{\IEEEauthorblockN{Robyn Ritchie\IEEEauthorrefmark{1},
Alexandre Leblanc\IEEEauthorrefmark{2}, Thomas Loughin\IEEEauthorrefmark{3}}
\\
\IEEEauthorblockA{\IEEEauthorrefmark{1}\IEEEauthorrefmark{3}Department of Statistics and Actuarial Science, Simon Fraser University, 8888 University Drive, Burnaby BC, Canada V5A1S6,\\ \IEEEauthorrefmark{2}Department of Statistics, University of Manitoba, 400 University Centre, Winnipeg MB, Canada R3T 2N2\\
Email: \IEEEauthorrefmark{1}robyn\_ritchie@sfu.ca,\IEEEauthorrefmark{2}alex.leblanc@umanitoba.ca,\IEEEauthorrefmark{3}thomas\_loughin@sfu.ca
}}

\maketitle
\begin{abstract}
    We introduce the most comprehensive publicly available datasets for mixed doubles curling, constructed from eleven top-level tournaments from the \hyperlink{https://curlit.com/results}{CurlIT} Results Booklets spanning 53 countries, 1,112 games, and nearly 70,000 recorded shots. While curling analytics has grown in recent years, mixed doubles remains under-served due to limited access to data. Using a combined text-scraping and image-processing pipeline, we extract and standardize detailed game- and shot-level information, including player statistics, hammer possession, Power Play usage, stone coordinates, and post-shot scoring states. We describe the data engineering workflow, highlight challenges in parsing historical records, and derive additional contextual features that enable rigorous strategic analysis. Using these datasets, we present initial insights into shot selection and success rates, scoring distributions, and team efficiencies, illustrating key differences between mixed doubles and traditional 4-player curling. We highlight various ways to analyze this type of data including from a shot-, end-, game- or team-level to display its versatilely. The resulting resources provide a foundation for advanced performance modeling, strategic evaluation, and future research in mixed doubles curling analytics, supporting broader analytical engagement with this rapidly growing discipline.
\end{abstract}


\newpage
\section{Introduction}

Over the past few decades, the quantity and quality of data in sports have increased dramatically. Technologies such as GPS sensors worn by players, optical tracking systems capturing player (x,y) coordinates multiple times per second, and biomechanical sensors embedded in equipment or apparel are now standard in many professional sports to assess movement, improve efficiency and consistency, and help prevent fatigue and injury (\cite{Seckin2023, Ronda2022}). However, many of these data sources are proprietary and unavailable for public use. In addition, publicly available data, such as play/event data, are typically restricted to major sports (baseball, basketball, football, hockey, soccer) that have the financial and infrastructural investments to collect them. Where such detailed event or tracking data are not accessible, researchers often must make do with coarse, game-level summaries.

Several researchers have worked to expand the availability of sports analytics data. For example, \cite{Thomas2017_nhlscrapr} compiled play-level data, shift charts, and shot locations in the National Hockey League (NHL), making these data available through the R package \texttt{nhlscrapR}. In American football, \cite{Yurko2019_football} developed \texttt{nflscrapR}, an R package that provides play-level data by parsing the National Football League’s official API. Beyond major sports, \cite{Zhang2024_billiards_sup} 
constructed a billiards dataset using the free software \hyperlink{https://www.kinovea.org/}{Kinovea} to collect the frames, turns and strikes from video footage. They then used the data to perform prediction, generate layouts and determine similar billiards layouts. In contrast, 4-player curling in general, and mixed doubles even more so, has received little investment in data infrastructure, limiting the scope of independent analytics in the sport.

Curling analytics as a whole remains in its infancy, with much of the existing research relying on game-level summaries to evaluate team ability and the strategic value of having the hammer. In a recent review, \cite{Zacharias2024} identified 38 articles focusing on curling strategy and tactics, but none addressing the mixed doubles format within their scope. Mixed doubles curling, often played in international competitions where tournaments rotate across countries and venues, presents unique logistical challenges for data collection. The temporary nature of venues complicates the installation of camera-based tracking systems, while GPS tracking of stones would require substantial investment and buy-in at the highest levels of the sport—resources typically unavailable outside elite tournaments.

Following the lead of prior work in other sports, we recognize the need for a publicly available mixed doubles curling dataset. Such data would enable researchers to investigate novel questions, foster independent analytics, and potentially provide a competitive edge within the sport.

In this paper, we describe the construction of a mixed doubles curling dataset at both the game and shot level and discuss the potential use cases. Starting from \hyperlink{https://curlit.com/results}{CurlIT}'s PDF tournament booklets, we extract tournament and game information, linescores, player and team statistics, and rock positions after each shot. We detail the text scraping, image processing, and post-processing methods used to clean the data and generate additional features. The resulting dataset spans 53 countries, 1,112 games, and 11 top level mixed doubles tournaments held between 2016 and 2025. The game-level dataset consists of one row per game and 167 variables describing the game-, team- and player-level statistics. The shot-level dataset contains 66,632 rows and 103 variables detailing the stone placements, the score and the number of stones in various zones of the play area following each shot.

The remainder of the paper is organized as follows. Section~\ref{Section:Curling_vs_MXD} reviews mixed doubles curling, highlighting its similarities and differences relative to standard curling. Section~\ref{Section:Data} outlines the data collection and feature engineering process. Section~\ref{Section:Results} presents elementary summaries intended to contrast the sports differences from standard curling and serve as a foundation for future research. Section~\ref{Section:Summary} concludes with directions for future research utilizing these datasets. 

\section{4-Player Curling vs. Mixed Doubles}\label{Section:Curling_vs_MXD}
\begin{wrapfigure}[18]{r}{0.4\linewidth}
    \includegraphics[width=0.98\linewidth]{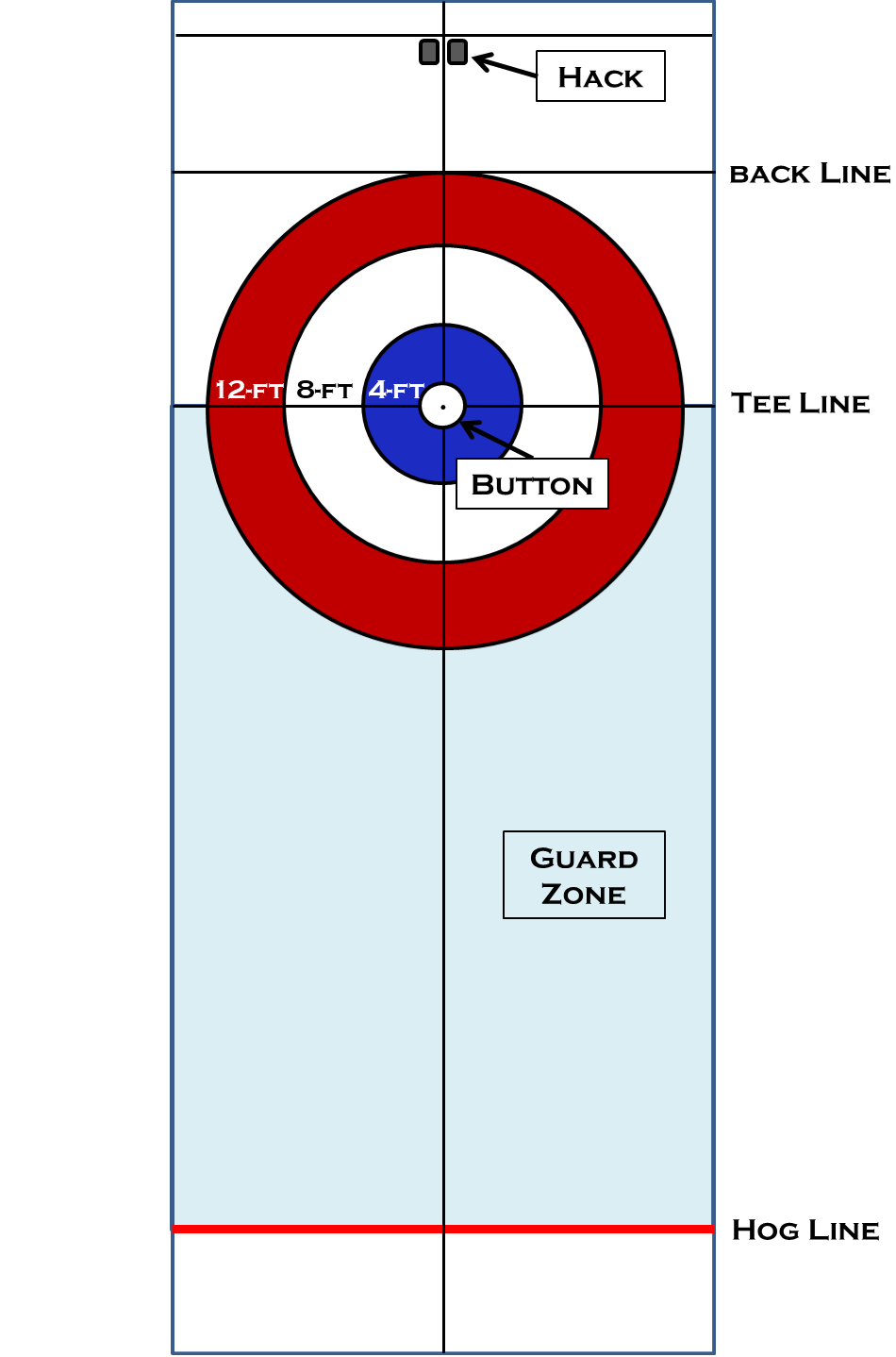} 
\caption{Layout of a sheet of ice in curling indicating the key features of the play area. Exact measurements can be found in the official Curling Canada Rule Book \citep{curling2022rulebook}.}
\label{fig:curling_sheet}
\end{wrapfigure}

Curling is a winter sport played on a sheet of ice approximately 150 feet (45.72 meters) long, in which teams slide heavy granite stones, or “rocks,” toward a target. The target, known as the “house,” consists of three concentric rings referred to by their diameters as the ``12-ft", ``8-ft", and ``4-ft" circles, shown in Figure~\ref{fig:curling_sheet} as red, white, and blue rings, respectively. A house is located at each end of the sheet. At the center of the house lies the “button,” a small white circle, whose exact center is known as the “pin”; a small hole drilled into the ice to allow precise distance measurements using a special device. Other key markings include 
the “center line,” which runs lengthwise through the sheet; the “tee line,” which runs across the sheet through the center of the house; and the “hog line,” the bold red line marking the start of the playing area which ends at the ``back line" labeled in Figure~\ref{fig:curling_sheet}.

A player attempts a shot by pushing from the “hack,” a foothold behind the house, and sliding a rock toward the opposite house while imparting rotation before release. This rotation causes the stone to follow a curved trajectory, or “curl”, which gives the sport its name.

A game consists of 8–10 intervals of play, called “ends,” in which two teams alternate throwing their rocks. Teams score when one of their rocks is in the house and closer to the pin than any of the opponent’s stones, earning one point for each stone in such a position. Accordingly, teams aim to place their rocks centrally in the house while displacing opposing stones. “Guard” rocks, stones positioned outside the house but within the playing area also known as the ``Guard Zone" (displayed in pale blue in Figure~\ref{fig:curling_sheet}), are often used to protect scoring stones from removal. The team with the final rock in an end is said to have the “hammer,” while the opposing team has the “first stone” possession. The hammer is a strategic advantage, as the final shot frequently determines scoring. If the team with the hammer scores, they forfeit the hammer in the next end. If the opposing team scores instead, which is known as a “steal,” the hammer is retained. When no points are scored, the end is said to be “blanked.”

Traditional curling is played between two teams of four players, each delivering two stones per end across the 8–10 ends, depending on the competition level. Standard curling has been a permanent part of the Olympic program since 1998 \citep{Olympics2025}. A newer variant, “mixed doubles curling,” has grown rapidly in popularity since its introduction at the World Mixed Doubles Curling Championships in 2008 \citep{FirstMixed} and its inclusion in the Olympics in 2018 \citep{OlympicsMixed}. Mixed doubles retains many of traditional curling’s fundamental mechanics, but key rule changes distinguish it from the 4-player format.

\begin{figure}
 	\centering
 	\includegraphics[width=0.22\textwidth]{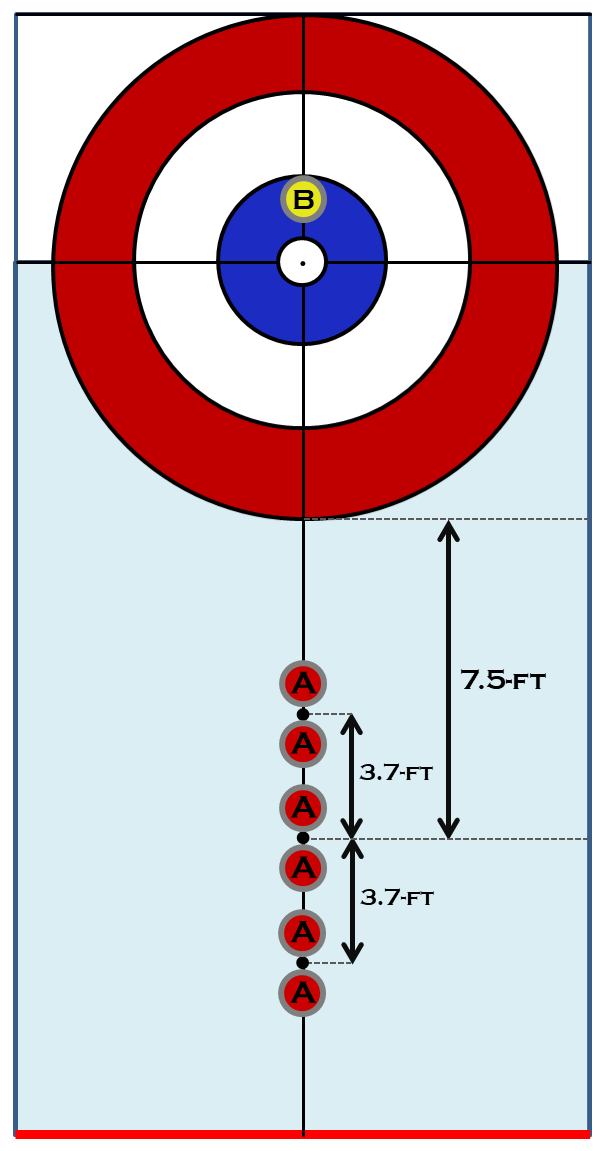}
 	\includegraphics[width=0.22\textwidth]{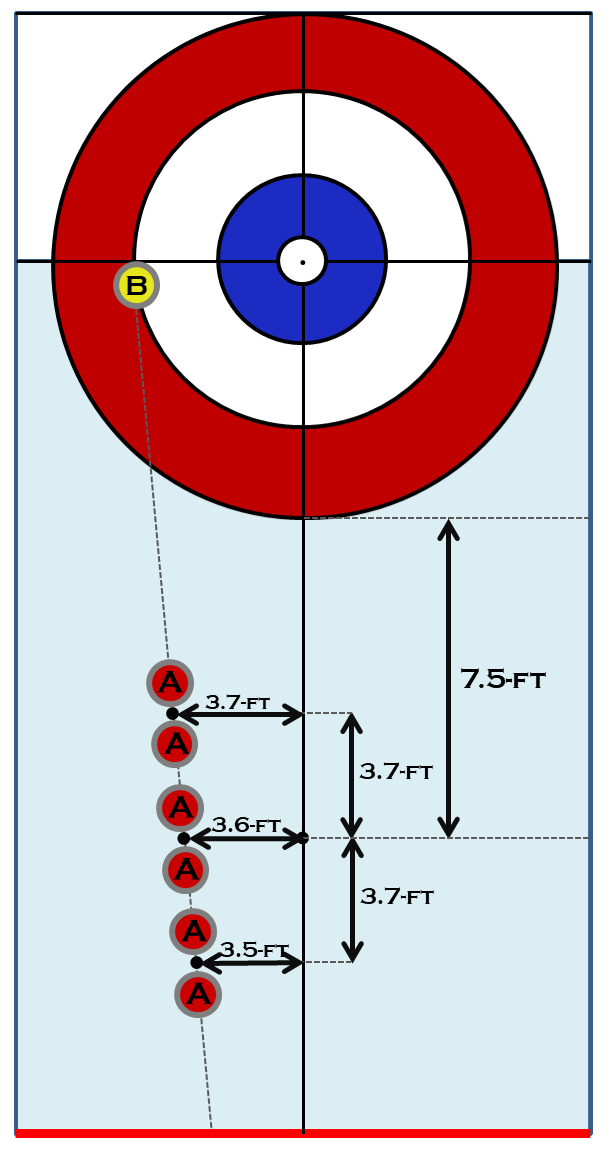}
 	\caption{Prepositioned Stones in mixed doubles for a standard end (left) and a Power Play end (right). The yellow team (position B) has the hammer in this end and the red team (position A) places their rock either in front or behind one of the three black ``guard dots" in the guard zone. Further measurement details can be found in the World Curling Rule Book \citep{CurlingRules}.
 	}
 	\label{fig:prepos_stones}
 \end{figure}

In mixed doubles, each team consists of two players, one male and one female, with games played over eight ends. Each team has six stones per end, one of which is prepositioned before play begins. The prepositioned stones are labeled as position A which is placed as a center guard and position B at the back of the 4-ft ring as shown in left image of Figure~\ref{fig:prepos_stones} by the red and yellow stones, respectively. The team whose stone is in position A delivers the first shot of the end and the team with position B delivers second and therefore has the hammer possession in the end. The team with position A can place their stone either in front or behind the game's black ``guard dot". The specific guard dot used in a game is determined before the game based on the ice conditions and will remain the same for the entirety of the game. Of the remaining five stones, one player delivers stones 1 and 5, while the other delivers stones 2, 3, and 4. 

Several rule modifications alter game dynamics relative to standard curling. In 4-player curling, the \textit{Free Guard Rule} stipulates that until the sixth stone of an end, a guard removed from play may be restored at the opponent’s discretion \citep{CurlingRules}. In mixed doubles, the rule is stricter: until the fourth thrown stone of an end, no rock may be removed from play, regardless of type (guard, in-house, prepositioned, or otherwise). If a removal occurs, all stones are reset by the non-offending team to their prior positions automatically, with no option for the opponent to accept the outcome. This increases the number of stones in play and raises the strategic importance of early shot placements.

Blanked ends are also treated differently. In standard curling, the hammer is retained if no points are scored, making blanking an important strategic option when teams prefer to preserve hammer advantage. In mixed doubles, however, a blanked end results in a \textit{change of control} as to where the prepositioned stones are placed in the next end. The team that takes stone position B, as in Figure~\ref{fig:prepos_stones}, gets the hammer. In nearly all cases, position B is selected and the hammer is taken. Consequently, blanking is both more difficult (given the number of stones in play) and not strategically advantageous. This change eliminates one of the central tactics in 4-player curling strategy, that has been investigated by several authors including \cite{Willoughby2005} and \cite{Clement2012}.

Another unique feature of mixed doubles is the Power Play. Once per game (excluding extra ends), the team with the decision on the prepositioned stones may call a Power Play at the start of an end. This shifts the prepositioned stones laterally: the hammer stone is placed intersecting the 8-ft and 12-ft rings just above the T-line, while the opposing guard is moved to the same side at a slight angle (Figure~\ref{fig:prepos_stones}, right). This configuration is intended to increase scoring potential by opening access to the button and reducing the likelihood of a steal. Teams may use the Power Play offensively, to attempt a large scoring end, or defensively, to minimize risk of giving up a steal.
  
While 4-player curling and mixed doubles share the same basic objectives and physical skills—such as sweeping, stone delivery, and shot placement—their strategic landscapes diverge considerably due to these rule changes. However, to date no research has examined whether strategies from traditional curling translate to mixed doubles. By making mixed doubles curling data publicly available, researchers can begin to study these differences systematically, gaining insight into how teams may need to adapt tactics for this distinct variant of the sport.

\section{Data collection}\label{Section:Data}
\begin{figure}
    \centering
    \includegraphics[width=0.7\linewidth]{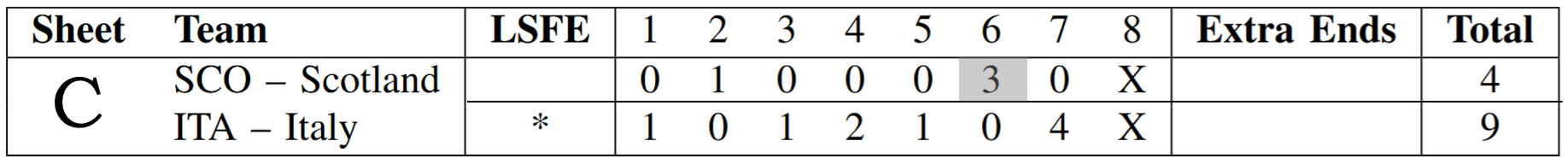}
    \caption{Linescore from the 2025 World Mixed Doubles Curling Championship for a game between Scotland and Italy. Last Stone First End (LSFE) is won prior to the game by Italy and thus they begin the game with hammer. The game is conceded prior to the final end and thus the score in that end is marked with `X'. Scotland's end 6 is highlighted in gray indicating they used their Power Play at this time. Italy did not use their Power Play.}
    \label{fig:mxd-linescore}
\end{figure}

Traditional curling analytics draws upon data recorded at varying levels of detail. The most common and accessible are game-level summaries, or linescores (shown in Figure~\ref{fig:mxd-linescore}), which track hammer possession and end-level results. These provide a progression of the game comparable to inning-by-inning summaries in baseball. Linescore data are widely available through sources such as the \hyperlink{https://stats.curling.io/}{Curling Canada Archive} and \hyperlink{www.worldcurling.org}{World Curling}, and have been frequently used in prior research due to their simplicity and public accessibility.

At a finer granularity, shot-level data record each delivered stone, along with contextual information and outcomes. \hyperlink{https://www.curlingzone.com/}{CurlingZone}, for instance, offers manually tracked shot by shot data for select events, though these resources are proprietary and not openly available. At the most detailed level, real-time performance tracking systems have been developed. \cite{Masui2015} introduced iCE (intelligent Curling Elicitator), a database platform that enables teams to log detailed shot information and player performance metrics for post-game review. While valuable for coaching and internal team analysis, these systems are likewise not publicly accessible.

Overall, access to curling data remains fragmented. Publicly available sources are generally limited to coarse, game-level summaries, whereas more detailed shot-level and performance data are proprietary. This restricted access has constrained the development of curling analytics and limited the scope of independent research—particularly in the context of mixed doubles.

\begin{figure}
    \centering
    \includegraphics[width=0.85\linewidth]{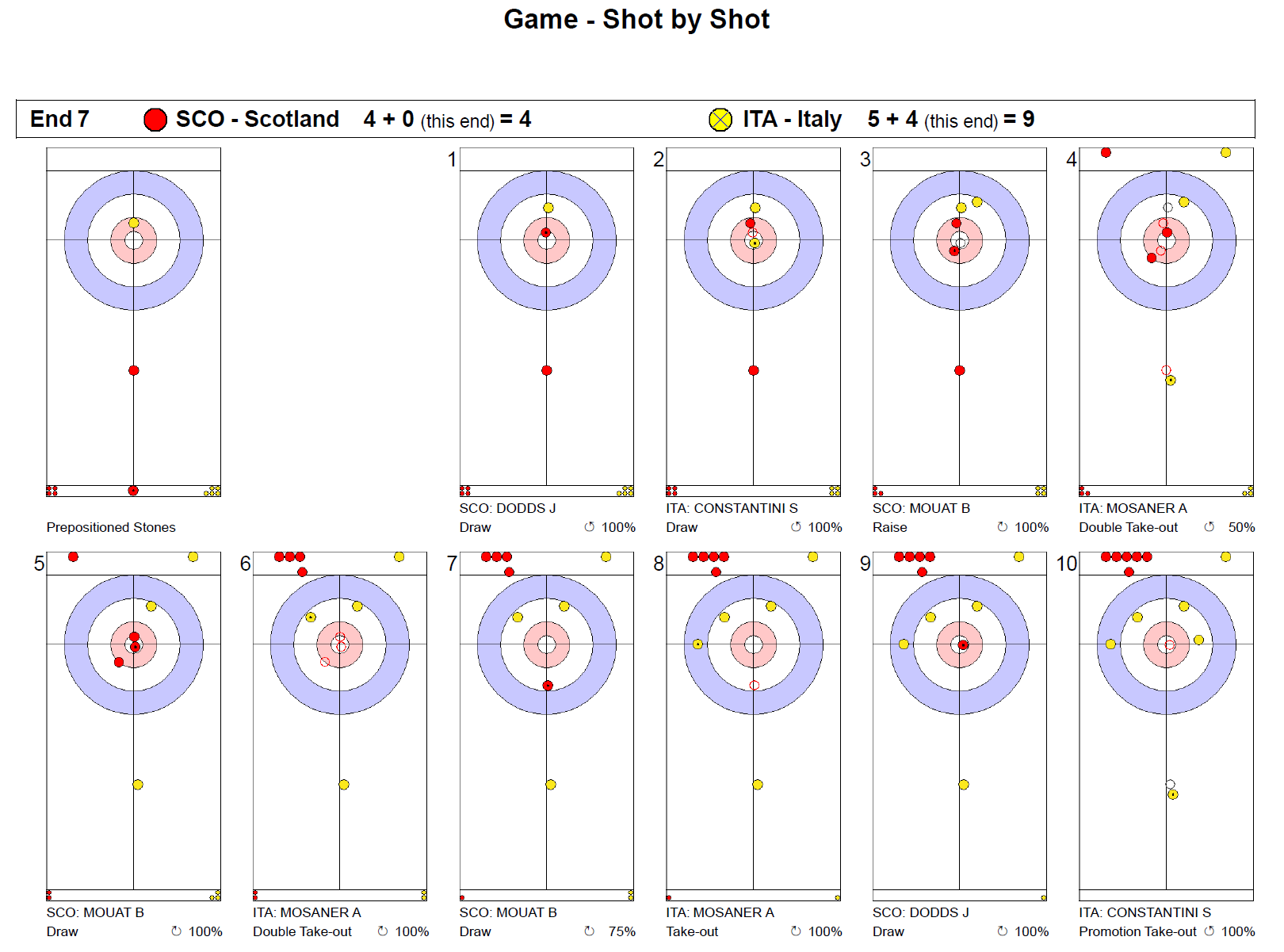}
    \caption{Game shot by shot frames from the  $7^{th}$ end between Scotland and Italy of the 2025 World Mixed Doubles Curling Championship \protect\hyperlink{https://curlit.com/results}{CurlIT} Results Booklet. The first image of the play area is the prepositioned stone set up prior to the first shot of the end.}
    \label{fig:mxd-shots}
\end{figure}

An important exception are the \hyperlink{https://curlit.com/results}{CurlIT} Results Books, which provide the most consistent and detailed publicly available curling data. Published for major competitions, these PDF booklets contain records dating back to 1924 for 4-player curling and 2013 for mixed doubles. Each Results Book includes team and player information, tournament schedules, standings, and game linescores. Beyond game outcomes, they report shot success percentages for both the game and tournament thus far, with breakdowns by shot type (draw, take-out, total) and by turn (clockwise, counter-clockwise, total). Most notably, the Results Books include graphical snapshots of the playing area after every shot, annotated with the shooting country, player, shot type, turn, and shot success percentage (Figure~\ref{fig:mxd-shots}). These detailed records make the \hyperlink{https://curlit.com/results}{CurlIT} Results Books the most comprehensive open resource for mixed doubles curling analytics.

To form our dataset, we focus on data from two types of tournaments: the World Mixed Doubles Curling Championships (WMDCC) from 2016 to 2025 (excluding 2020) and the Olympic Winter Games (OWG) in 2018 and 2022. These tournaments were selected because the \hyperlink{https://curlit.com/results}{CurlIT} Results Books follow a consistent reporting format across years, and we believe the level of competition is relatively comparable. 

Data collection was conducted in four stages using both Python and R. The primary step involved text scraping the \hyperlink{https://curlit.com/results}{CurlIT} Results Books to extract structured data on teams, players, games, shots, and standings. Next, we retrieved and processed all shot by shot images included in the Results Books. Image processing was used to identify stone positions, stone colours, and the shooter’s stone for each shot. The extracted data then underwent cleaning and validation to ensure completeness and consistency. Country and player names were standardized across tournaments, and missing information (e.g., absent shot images, unidentified shooters, or unspecified shot types) was corrected wherever possible. Finally, a post-processing stage derived additional features critical to mixed doubles curling analysis. These include hammer possession for each end, identification of Power Play usage (end and team), and whether Power Plays remained available. Using standard curling terminology, we also classified stone positions into defined zones based on their final resting locations, enabling systematic analysis of shot outcomes.

\subsection{Text Scraping}
Text scraping was used to convert the unstructured tournament records in the \hyperlink{https://curlit.com/results}{CurlIT} Results Books into structured data suitable for analysis. The process involved systematically scanning page content, identifying patterns (e.g., keywords, abbreviations, and formatting conventions), and extracting relevant entities such as teams, players, standings, and shot outcomes. Country codes and player names posed the greatest challenges for automated parsing due to occasional formatting irregularities, while numerical fields such as scores, standings, and shot success percentages were generally extracted with minimal intervention.

Not all pages from the Results Books were scraped; instead, we focused on those providing unique and essential information. Specifically:
\begin{itemize}
    \item \textbf{Title Page, Medalists, and Final Standings}: Tournament metadata (location, dates), medal results, and final rankings.
    \item \textbf{Game Results}:  Linescores, Last Stone Draw outcomes, timeouts, time remaining, and Power Play usage. In addition, player- and team-level shooting percentages for the game as well as the tournament thus far are provided along with shot success analysis providing a detailed breakdown of these percentages by stone rotation and grouped by draw or take-out and a combined total,
    \item \textbf{Shot Results}: When provided, pages include detailed records of each shot, including shooter, shot type, rotation, and execution rating.
\end{itemize}
While most tournaments follow consistent formatting, occasional discrepancies—such as missing data, ambiguous player identifiers, or format shifts—required manual corrections to maintain comparability across years. Detailed descriptions of these interventions are provided in the Supplementary Material (Section~\ref{Sup:Textscraping}).

\subsection{Image Extraction}
Shot by shot images were extracted from the \hyperlink{https://curlit.com/results}{CurlIT} Results Books using the \texttt{fitz} package from \texttt{PyMuPDF} in Python. For each shot by shot page, we identified all embedded images with approximate dimensions of 300 × 600 pixels and saved them as shot images. The standard layout, as seen in Figure~\ref{fig:mxd-shots}, consists of one image showing the prepositioned stones, followed by sequential images of shots 1–4 on the first row and shots 5–10 on the second row.

While this procedure was largely automated, several exceptions required manual intervention. For example, the 2016 WMDCC did not include prepositioned stone images. To address this, we manually reconstructed these images by examining the first shot of each end and cross-referencing stone placement, colours, and sheet orientation with the 2017 tournament format. In later tournaments, occasional discrepancies occurred, most commonly when the prepositioned stones image was mistakenly duplicated from the first shot image. In such cases, the incorrect images were replaced with corrected versions sourced from other ends or games with the appropriate configuration.

A complete list of manual corrections, excluding those applied to the 2016 WMDCC, is provided in Supplementary Material (Section~\ref{Sup:Image_Issues}, Table~\ref{tab:Preposition_Stone_Fixes}).

\subsection{Image Processing}
The shot by shot images extracted from the Results Books required additional processing to standardize orientation, identify stone positions, classify stone colours, and detect the shooter. This step was essential to convert static visual records into structured data suitable for quantitative analysis. The overall workflow was designed to ensure that each shot image could be consistently compared across games, ends, and tournaments despite minor formatting differences between years.

\begin{figure}
    \begin{center}
        \includegraphics[width=0.85\linewidth]{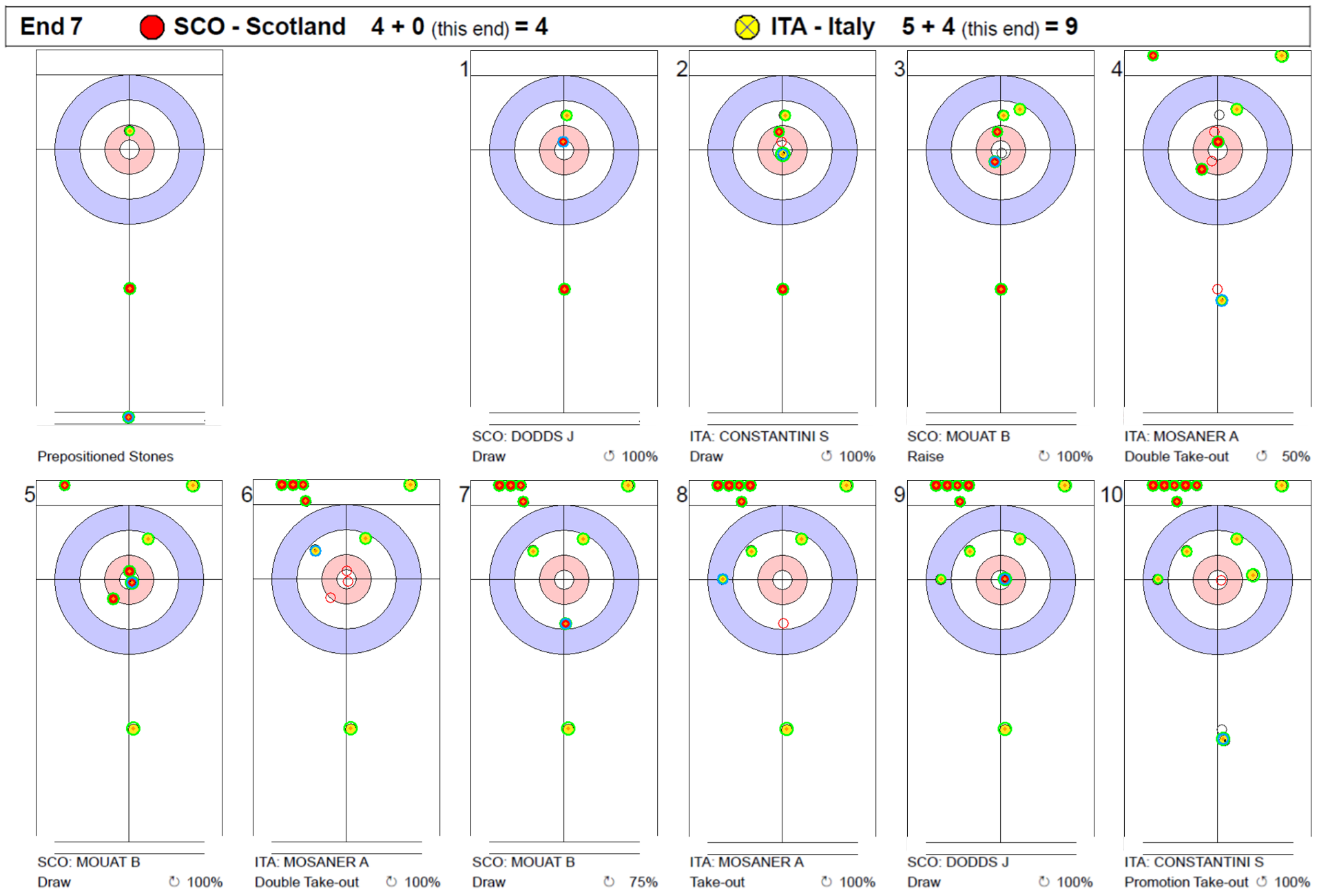}
        \caption{Shot by shot images from Figure \ref{fig:mxd-shots} after processing to detect stones, their colours and the shooter location.}
        \label{fig:shots_detected}
    \end{center}
\end{figure}

The processing pipeline proceeded as follows:
\begin{enumerate}
    \item \textbf{Orientation}: Images were standardized by flipping when necessary so the house always appeared in a consistent position.
    \item \textbf{Stone Detection}: Stones were identified using circle-detection methods, with unused stones in the lower corners masked out.
    \item \textbf{Colour Classification}: Stones were classified as red or yellow based on pixel counts, with shooter colour alternating with each shot.
    \item \textbf{Shooter Identification}: The shooter was identified using tournament-specific markings (a black border in earlier years; a central black dot in later years).
    \item \textbf{Sequence Validation}: When image extraction order was inconsistent, stone coordinates were checked and corrected to restore chronological order.
\end{enumerate}
Figure~\ref{fig:shots_detected} illustrates the processed images from Figure \ref{fig:mxd-shots} of the $7^{th}$ end in the 2025 WMDCC game between Scotland (red stones) and Italy (yellow stones). Comparing to the original images, the processed images are rotated, lower corners whited out (to avoid detecting unused stones), stones are detected (bright green circles), stone colours are identified, the shooter is highlighted (blue circle), and precise stone coordinates are marked (orange dots).

\subsection{Post-Processing}
After extracting and processing the shot images, we conducted a post-processing step in R and Python to create comprehensive datasets describing both game- and shot-level details. This stage involved additional cleaning, feature creation, and the integration of contextual game information needed for subsequent analysis.

\begin{figure}
\begin{center}\includegraphics[width=0.4\linewidth]{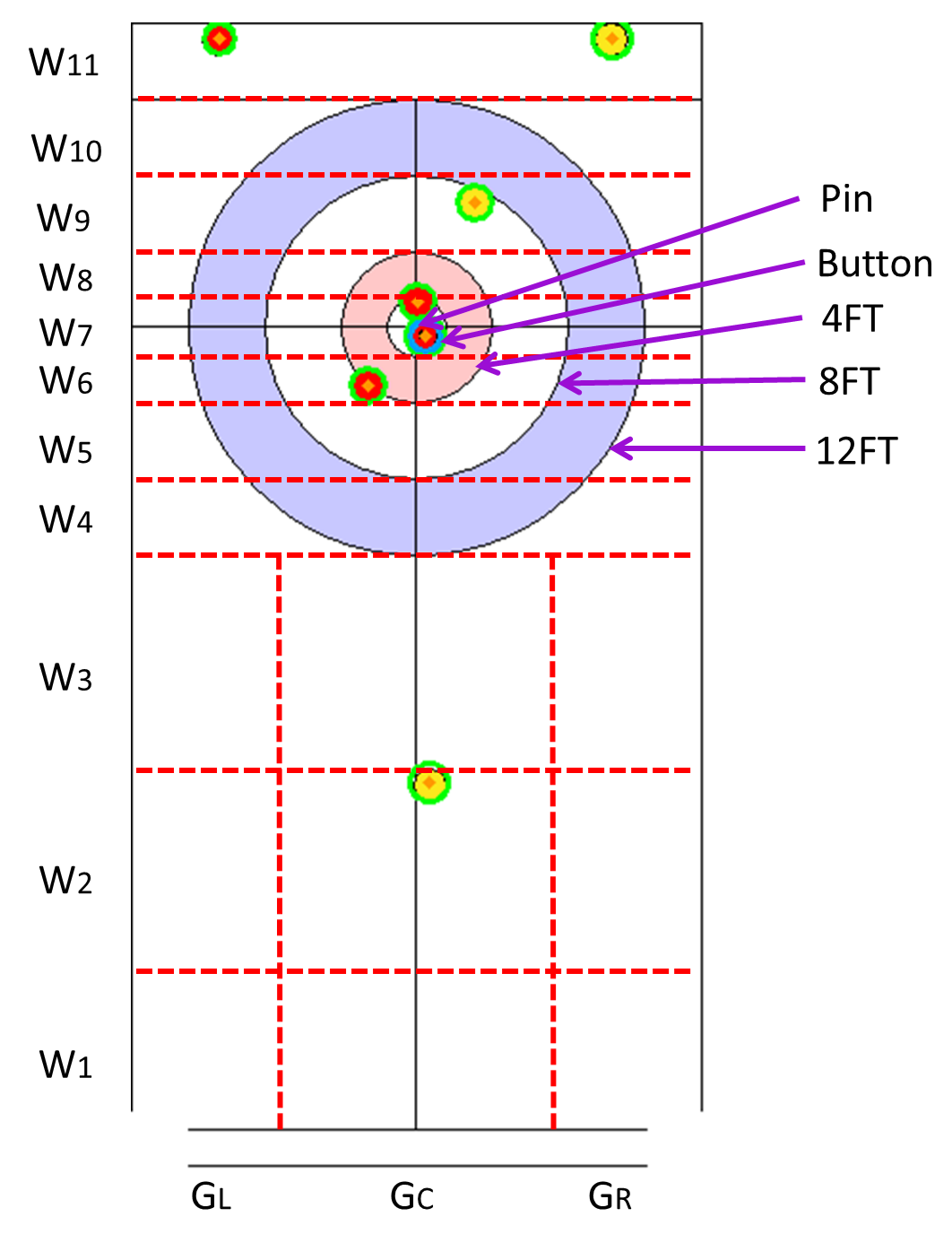}
        \caption{Map for determining the number of stones within each zone and ring of the house. These can be used as features in models to describe the situation of the current shot.}
        \label{fig:weights}
    \end{center}
\end{figure}

The main post-processing workflow included:
\begin{enumerate}
    \item \textbf{Cleaning and Standardization}
    \begin{itemize}
        \item Tournament standings were standardized by converting medal labels (Gold, Silver, Bronze) into numerical ranks (1, 2, 3) to align with the remaining rankings.
        \item Thrower names were cross-checked with the official Entry List to ensure each shot was correctly assigned to the shooting country, labeled as the variable ``Rock Country" in Figure~\ref{fig:shots_data}.
    \end{itemize}
    \item \textbf{Hammer and Power Play Features}
    \begin{itemize}
        \item The country delivering the second shot of each end was identified as having the hammer and given the variable name ``Hammer End" in Figure~\ref{fig:shots_data}.
        \item Power Plays were validated by checking that the average X-coordinate of prepositioned stones was offset from center ice. From this, we determined whether the Power Play was set to the left or right side of the sheet.
        \item New variables were added to indicate Power Play availability at the start of each end, ensuring consistency with hammer rules.
        \item Hammer possession in the first end was used to confirm the Last Stone First End (LSFE) winner, as listed in the game results.
    \end{itemize}
    \item \textbf{Team Stone Colour Assignment}
    \begin{itemize}
        \item By comparing the first and second shot colours with hammer possession, we determined each team’s stone colours for the game.
    \end{itemize}
    \item \textbf{Stone Zone Classification}
    \begin{itemize}
        \item Moving back to Python, stones were categorized into zones based on their final resting location (Figure~\ref{fig:weights}). Zones ranged from weight based (W1 for high guards to W11 for beyond play area) to house rings (Button, ``4-ft", ``8-ft", ``12-ft"). The red stone in the ``8-ft" ring of Figure~\ref{fig:weights} would get a value of 1 for both the ``RedW6" and ``Red8FT" columns and the score after the shot would be ``Red\_Score" is 3 and  ``Yellow\_Score" is 0.
        \item Stones in front the house were further classified as left (GL), center (GC), or right (GR) guards.
        \item After each shot, scoring was computed by finding the stone closest to the button and counting all stones of the same colour in the house and closer than the nearest opponent stone.
    \end{itemize}
\end{enumerate}

The final datasets included both \textit{game}- and \textit{shot}-specific tables, with column descriptions provided in the Supplementary Material (Section~\ref{Sup:columns}). For compactness, we use \textbf{\{\}} notation to describe similar columns. For example, the column \textbf{T\{1,2\}\{F,M\}\{Draws,TO,TotalGame\}\{Clock,Counter\}Num} encodes team (1 or 2), gender (female or male), shot type (draws, take-outs (TO), or game totals), and rotation (clockwise or counter-clockwise). A specific selection such as T1FTOCounterNum would represent the number of counter-clockwise take-outs attempted by the female player on Team 1 during the full match.

Table~\ref{tab:Table_Stats} provides a summary of the mixed doubles curling datasets. The game-level dataset includes 1,112 games played by 274 players from 53 countries. The shot-level dataset includes 804 games as not all games include shot by shot details. This dataset includes data from 40 teams, 207 players and results in over 60,000 recorded shots. It also captures detailed end-play structures, including 4,683 regular ends, 1,352 Power Play ends, and 61 overtime ends. This breadth and depth of data enables robust statistical analyses of player behavior, team strategies, and shot success across a diverse international field.
\begin{table}
 	\centering
 	\begin{tabular}{lcccccccc}
 		\hline
 		Dataset & Years & Countries & Players & Games & Reg Ends & PP Ends & OT Ends & Shots\\
 		\hline
        Games & 2016-2025 (excl. 2020) & 53 & 274 & 1,112 & - & - & - & -\\
        Shots & 2016-2025 (excl. 2020) & 40 & 207 & 804 & 4,683 & 1352 & 61 & 60,536\\
            \hline\\
 	\end{tabular}
 	\caption{Summarizing numbers for the data included in the mixed doubles datasets.
    }
 	\label{tab:Table_Stats}
 \end{table}

\subsection{Accessing the Data}
The datasets described in this section are publicly available in the \texttt{CurlR} repository. The data can be accessed though GitHub at \hyperlink{https://github.com/ritchi12/CurlR}{CurlR} by using the commands found in Listing~\ref{list:data_access}. This repository provides comprehensive data tables and accompanying documentation to facilitate reproducible research and further analysis in the field of sports analytics.

~\begin{lstlisting}[language=R, caption={Install CurlR using devtools and load data}, label={list:data_access}]
devtools::install_github("ritchi12/CurlR")
library(CurlR)
data("mxd_games")
data("mxd_shots")
\end{lstlisting}
 
\section{Results and Insights from the Mixed Doubles Curling Dataset}\label{Section:Results}
\begin{figure}
    \begin{center}
    \includegraphics[width=1\linewidth]{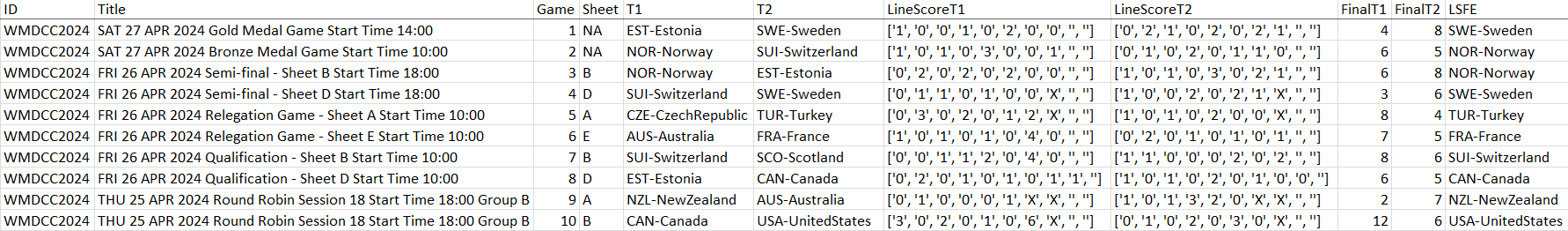}
        \hspace{0.5mm}
        \caption{Sample game-level data from the 2025 World Mixed Doubles Curling Championship games showing the teams, linescores, final scores, and LSFE.}
        \label{fig:game_data}
    \end{center}
\end{figure}

\begin{figure}
    \begin{center}
        \includegraphics[width=1\linewidth]{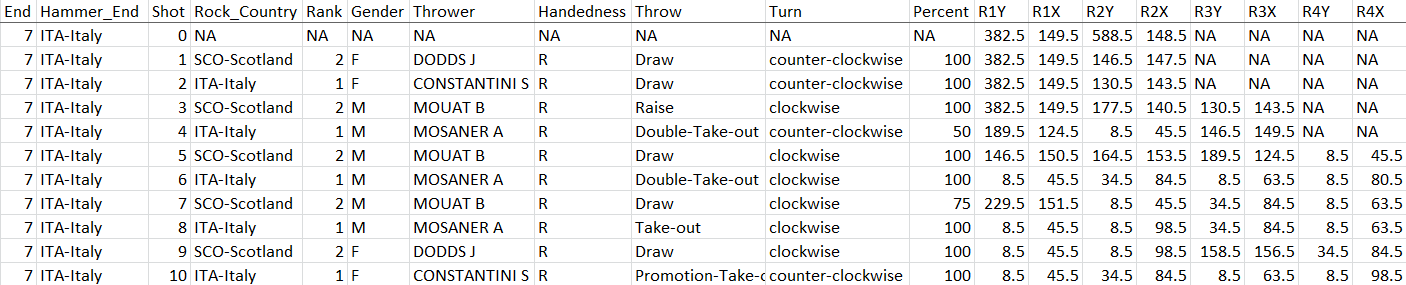}
        \hspace{0.5mm}
        \caption{Sample shot-level data for the $7^{th}$ end of the 2025 World Mixed Doubles Curling Championship gold medal game between Scotland and Italy. Details include the end, team with hammer in the end, shot number, country throwing the rock, rank in the tournament, gender, name and handedness of the thrower, throw type, turn and percent grading of the shot and some of the (X,Y) coordinates of the red stones.}
        \label{fig:shots_data}
    \end{center}
\end{figure}

The curated datasets provide a rich foundation for quantitative analysis of mixed doubles curling, comprising 268 detailed columns across game- and shot-level data. Figures~\ref{fig:game_data} and \ref{fig:shots_data} present representative samples from the game and shot datasets, respectively. Shot number 0 denotes the prepositioned stones. With detailed shot- and game-level datasets, the data enable multifaceted analyses of game strategy and performance, allowing for extensions to analyze ends and teams. 

\subsection{Shot-Level Analysis}
\begin{figure}
    \centering
        \includegraphics[width=0.45\linewidth]{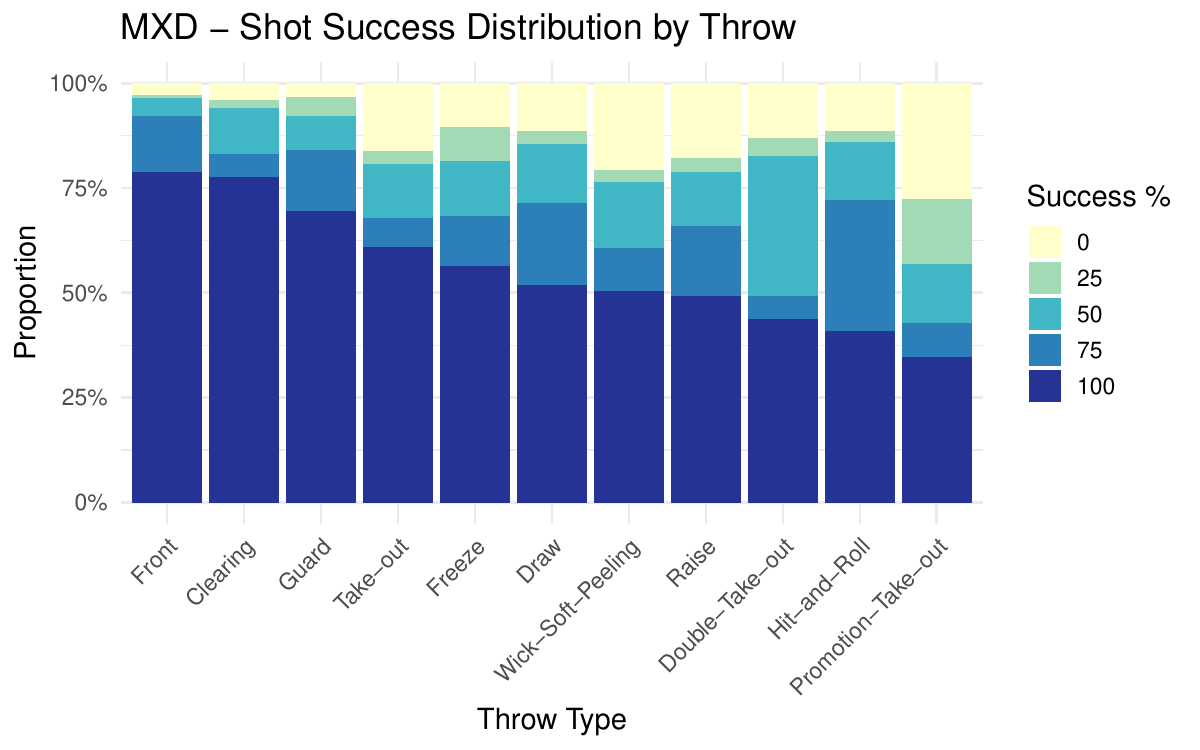}
        \includegraphics[width=0.45\linewidth]{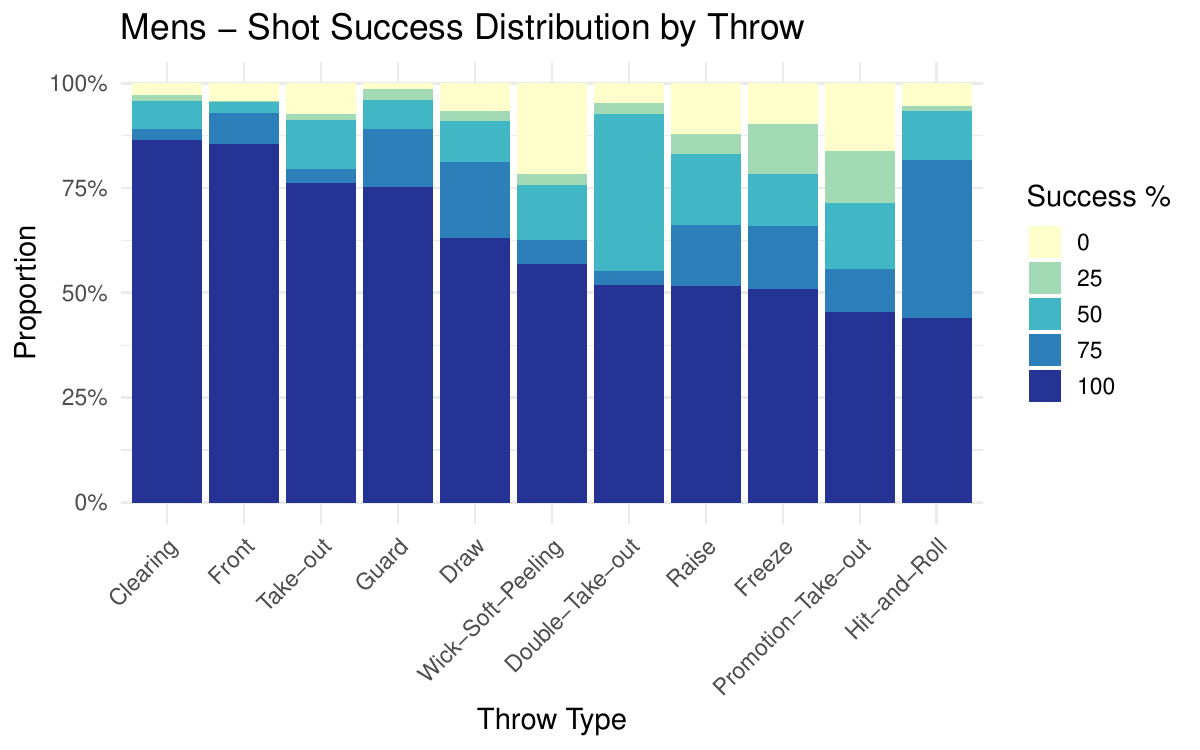}
        \caption{Bar chart displaying the interaction of columns Throw and Percent in the shot-level dataset. We compared the mixed doubles, MXD, tournaments (left) to the men's (right).}
        \label{fig:shot_succ}
\end{figure}

Each shot is graded on a discrete scale—0, 25, 50, 75, or 100—by trained personnel capturing data live during events. These shot grades reflect execution success relative to the intended outcome. However, some subjectivity or uncertainty is inherent, as players may not always explicitly announce their shot intentions or may adapt mid-execution. Comparing mixed doubles success proportions to the World Men's Curling Championships (WMCC2019, WMCC2021, WMCC2022, WMCC2023, WMCC2024) in Figure~\ref{fig:shot_succ}, front, clearing, and guard shots have the highest average success percentages in mixed doubles. The fraction of 100\% scores received in the men's game (right panel) are higher across all shot types with the largest difference appearing in take-outs. Overall, the most successful shots are those that typically have less risk.

\begin{figure}
    \centering
    \includegraphics[width=0.5\linewidth]{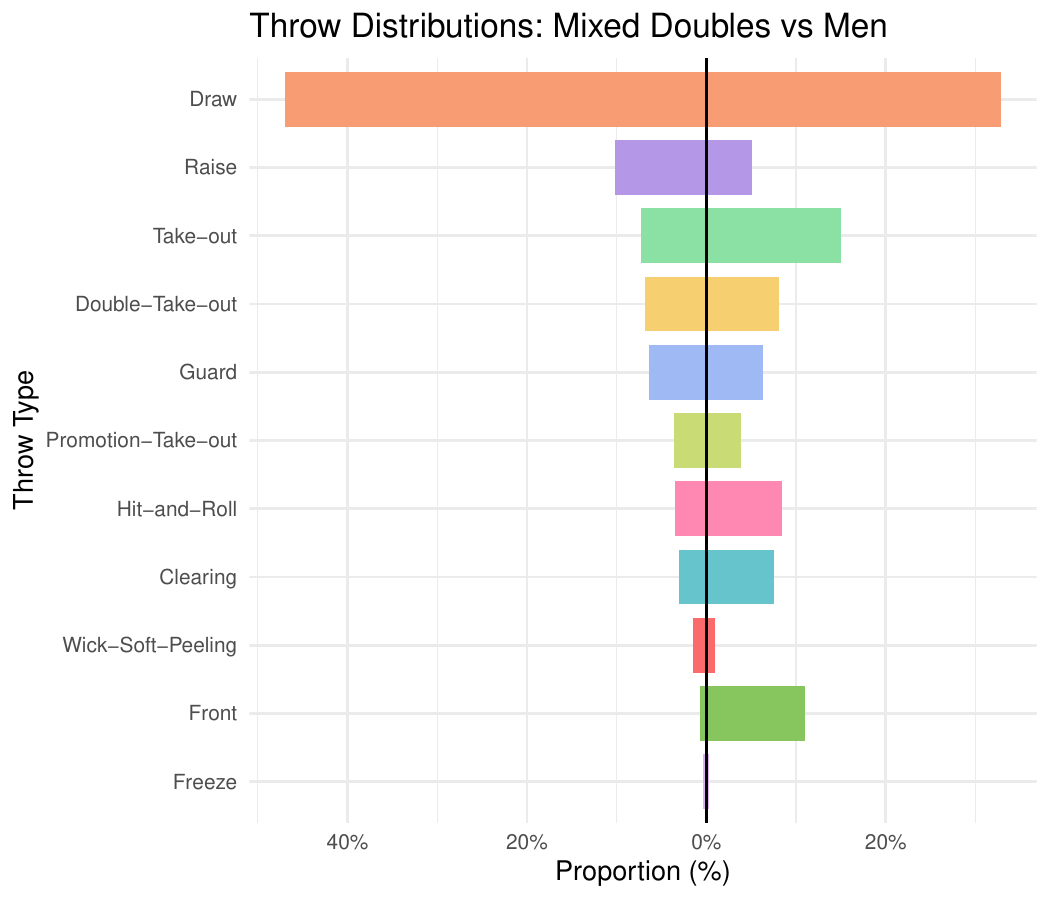}
        \caption{Back-to-back bar chart displaying the proportion of each type of Throw attempted in the shot-level dataset for the mixed double's (left) compared to men's (right) tournaments.}
        \label{fig:bar_plots}
\end{figure}

Looking at the proportion of throws attempted in Figure~\ref{fig:bar_plots}, draws dominate the plot for both the mixed doubles and men's games. For the men, take-outs, hit-and-rolls, clearing and fronts have much larger shares of the action. The types of throws attempted in the two versions of curling are drastically different further enforcing that any analytics should take these aspects into consideration.

\subsection{End-Level Analysis}\label{EndAnalysis}
The scoring behavior in mixed doubles curling differs significantly from traditional 4-player curling. While standard curling scoring distributions are approximately symmetric and centered around scoring 1 point with the hammer (figures available in \cite{Brenzel2019}), mixed doubles has a gap in the distribution because scoring zero is both difficult and not usually beneficial to the team with hammer. This leads to a fundamentally different scoring distribution in the two variations of the sport.

\begin{figure}
    \begin{center}
        \includegraphics[width=1\linewidth]{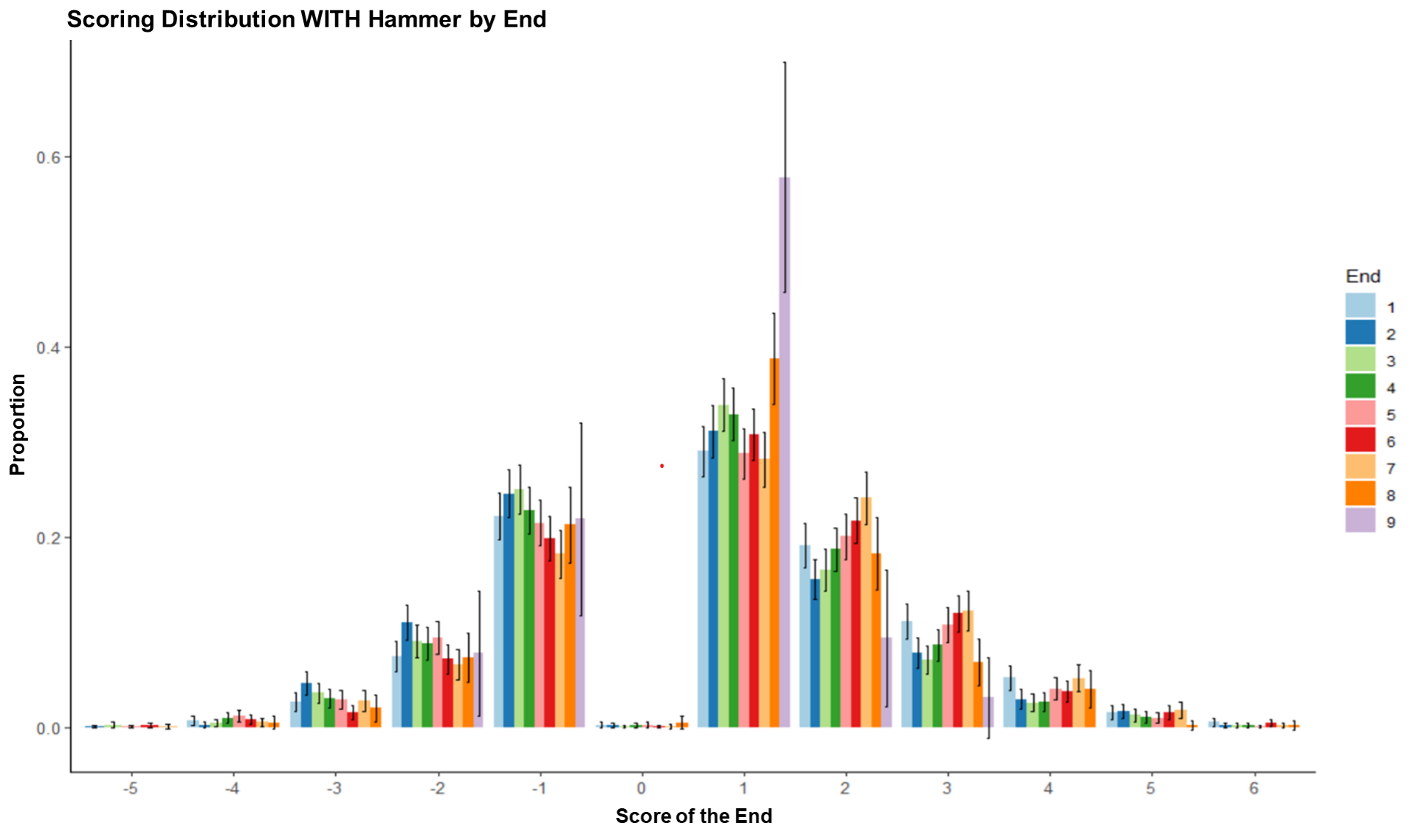}
        \caption{Scoring distribution of the hammer team by end.}
        \label{fig:mx_scoring_dist_end}
    \end{center}
\end{figure}

\begin{figure}
    \begin{center}
        \includegraphics[width=1\linewidth]{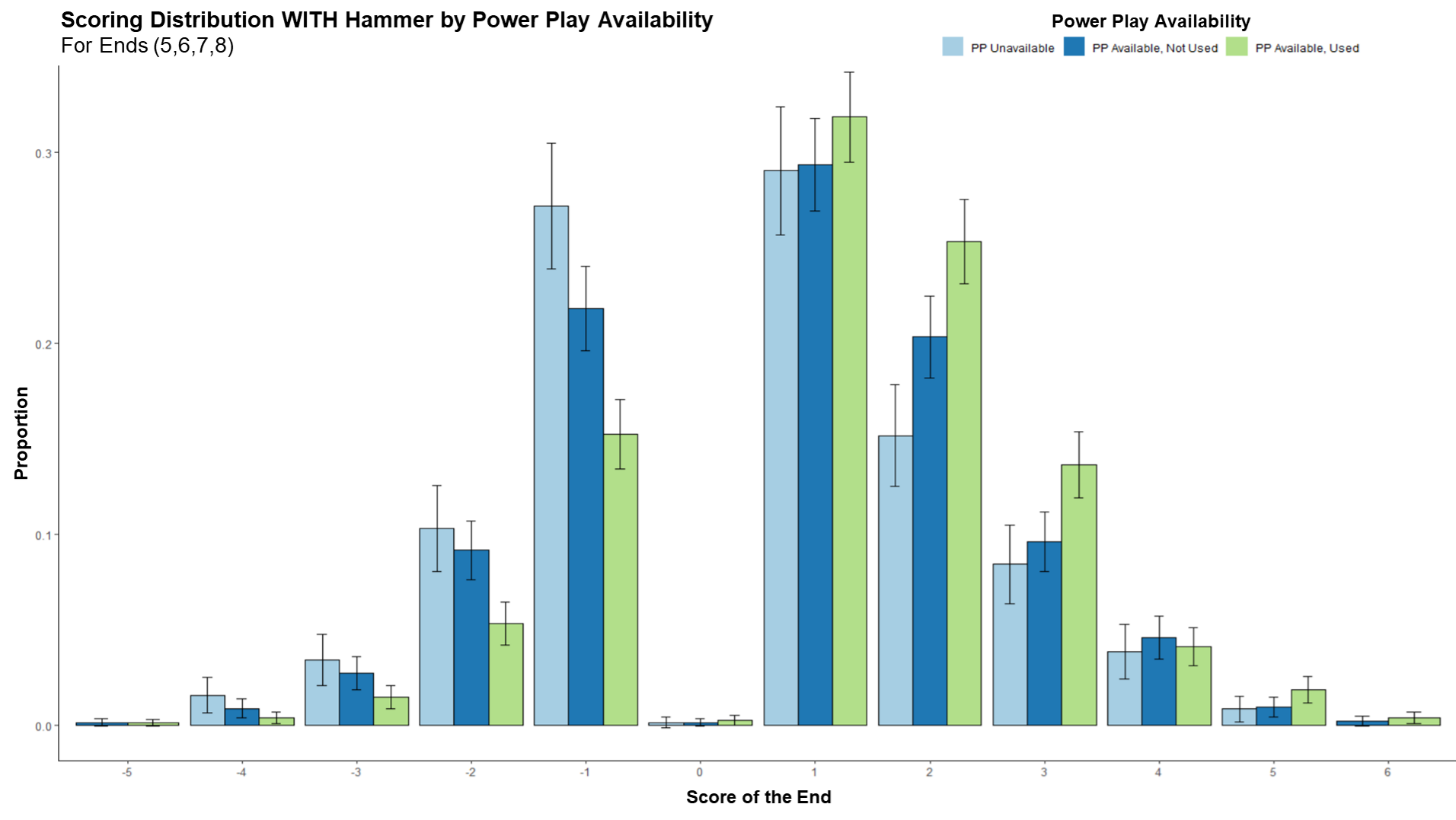}
        \caption{Scoring distribution by Power Play availability of the hammer team in ends 5 through 8.}
        \label{fig:mx_scoring_dist_pp}
    \end{center}
\end{figure}

As illustrated in Figure~\ref{fig:mx_scoring_dist_end}, the distribution of points scored across ends is relatively consistent, with a notable shift in the 9th end, when there is a much higher frequency of scoring one point with hammer, because that is all they need to win. When it comes to scoring more than one point, end 7 has the highest frequency out of all ends. This may be attributed to strategic use of the Power Play, typically deployed in later ends (5–8). Figure~\ref{fig:mx_scoring_dist_pp} further dis-aggregates scoring outcomes based on Power Play status:
\begin{itemize}
    \item PP Available \& Used,
    \item PP Available \& Not Used, and
    \item PP Unavailable.
\end{itemize}
Teams using the Power Play tend to increase their likelihood of scoring 2 or 3 points while reducing the chances of conceding a steal. When the Power Play is unavailable, the likelihood of conceding a single point rises. This effect may be influenced by weaker teams using their Power Play earlier because they are trailing, or by leading teams being more willing to concede a point late in the game. Figures~\ref{fig:mx_scoring_dist_b4diff_end} and \ref{fig:mx_scoring_dist_pp_end} in the Supplementary Material (Section~\ref{Sup:Scoring}) extend this analysis by overlaying Power Play availability with end number and score differential, providing a multi-dimensional view of scoring tendencies.

\subsection{Game-Level Analysis}
The game-level dataset contains high-level summaries of performance for each team across all ends, capturing aggregate statistics such as draw, take-out and total success percentages. These features allow for longitudinal team performance evaluation and broader tournament-level insights.

\begin{figure}
    \begin{center}
        \includegraphics[width=0.8\linewidth]{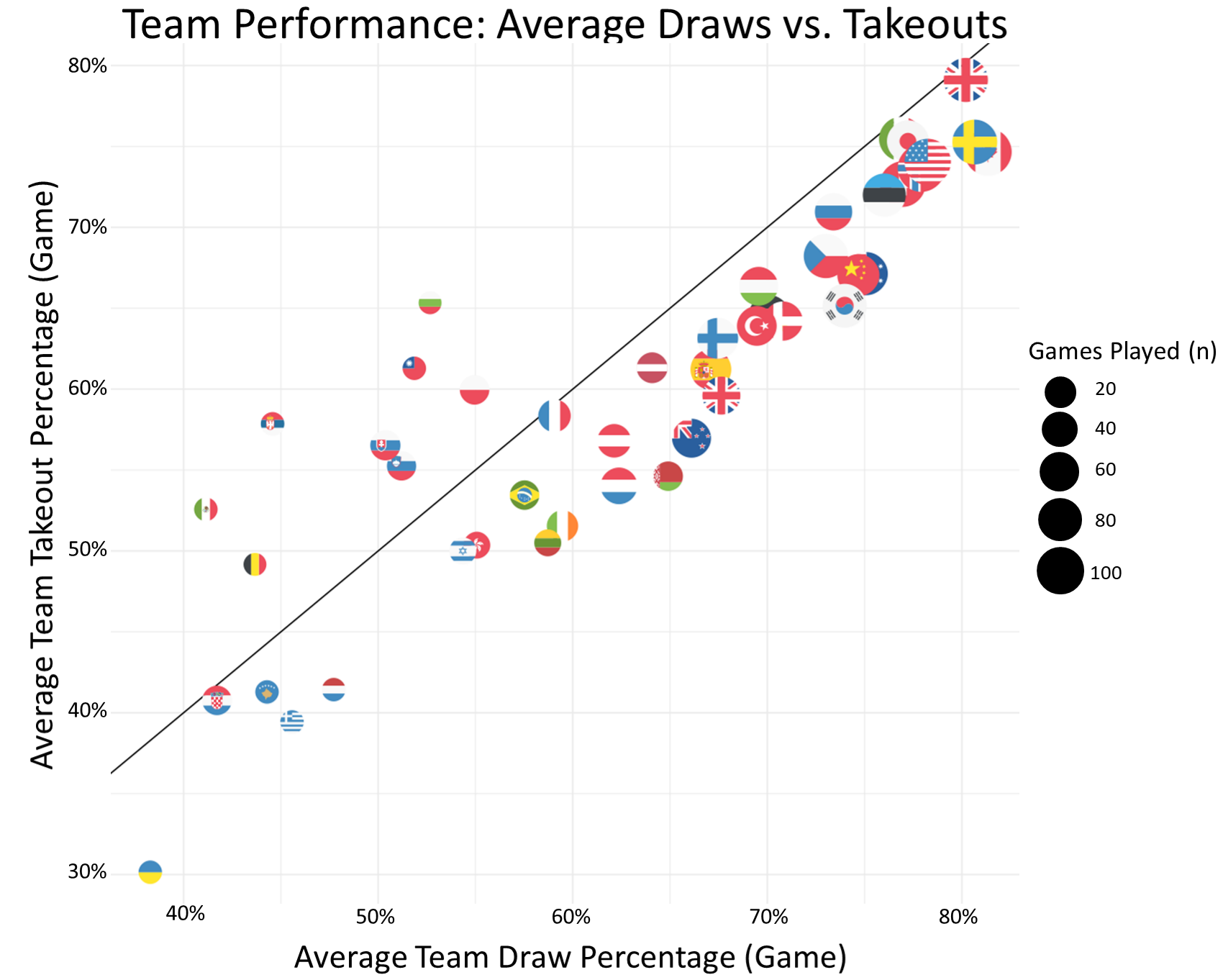}
        \caption{Averaged draw vs. take-out percentages for teams  across games in the game-level dataset. Sizes of each point corresponds to the number of games each team played.}
        \label{fig:flags_draws_to}
    \end{center}
\end{figure}

Figure~\ref{fig:flags_draws_to} visualizes the relationship between a team’s average draw and take-out success percentages over the games in the dataset, based on the \textbf{T\{1,2\}DrawsGamePercent} and \textbf{T\{1,2\}TOGamePercent} columns. Each point represents a national team, and its size reflects the number of games played. The strongest overall performers include Great Britain, Canada, Sweden, Italy, Japan, USA, Norway, Estonia, and Switzerland, all averaging above 75\% in both draw and take-out success. These results suggest that consistent shot-making ability in both draws and take-outs strategies is strongly related to team success.

Most top-performing teams exhibit a balanced proficiency in both draws and take-outs, clustering near the top-right quadrant of the plot. However, the majority of teams have a superior performance in draws compared to take-outs. The scoring in curling is on a 0-100 scale (taking only steps of 25) and describes the success of the shot. A draw missed by half a foot may still get a score of 75, but a take-out missed by an inch can cause irreparable damage to an end and get a score of 0. A new scoring system that incorporates the level of difficulty of a shot could help to better evaluate team and player performances. 

The game-level dataset also captures nuanced strategic behavior, such as Power Play selection timing and hammer vs. first stone possession. These can be integrated with end-level and shot-level data to model more complex dynamics, such as momentum shifts, comeback likelihoods, or optimal Power Play usage patterns across games.

\subsection{Team-Level Analysis}
Since athlete participation can vary from year to year, analyses are conducted primarily at the country level rather than at the individual player or team level, thereby increasing sample sizes and ensuring comparability across tournaments.

To assess team performance, we adopt the efficiency metrics proposed by \cite{Palmer2019} and adapt them for mixed doubles as follows,
\begin{itemize}
    \item Hammer Efficiency (HE): Proportion of ends with hammer where a team scores 2 or more points.
    \item Force Efficiency (FE): Proportion of ends without hammer where a team holds the opponent to one or zero points. Unlike traditional curling, we include blanked ends in this definition as the change in hammer possession benefits the team without hammer, although these rarely occur in mixed doubles.
    \item Steal Efficiency (SE): Proportion of ends where a team scores at least one point without hammer.
\end{itemize}
These efficiency metrics are presented in Table~\ref{tab:hammer_efficiency}, both for standard ends and Power Play ends. Notably:
\begin{itemize}
    \item Canada leads in standard Hammer Efficiency (40.73\%).
    \item Japan achieves the highest Force Efficiency (40.11\%).
    \item Ireland tops Steal Efficiency at 51.28\%.
    \item In Power Play situations, Israel dominates Hammer Efficiency, while Austria and France lead in Force and Steal Efficiencies, respectively, although all of these teams have played a relatively small number of games which as indicated in the parentheses in Table~\ref{tab:hammer_efficiency} next to the proportions.
\end{itemize}
\begin{table}
 	\centering
 	\begin{tabular}{lcccccc}
 		\hline
 		Country & HE & PP-HE & FE & PP-FE & SE & PP-SE \\
 		\hline
CAN-Canada & \textbf{0.4073} \tiny{\textbf{(275)}} & 0.4407 \tiny{(59)} & 0.3732 \tiny{(284)} & 0.3647 \tiny{(85)} & 0.4648 \tiny{(284)} & 0.3059 \tiny{(85)}\\
SWE-Sweden & 0.3704 \tiny{(216)} & 0.4561 \tiny{(57)} & 0.3550 \tiny{(231)} & 0.3099 \tiny{(71)} & 0.4372 \tiny{(231)} & 0.3662 \tiny{(71)}\\
SCO-Scotland & 0.3702 \tiny{(235)} & 0.4507 \tiny{(71)} & 0.3577 \tiny{(246)} & 0.3239 \tiny{(71)} & 0.4350 \tiny{(246)} & 0.3380 \tiny{(71)} \\
SUI-Switzerland & 0.3402 \tiny{(244)} & 0.3385 \tiny{(65)} & 0.3252 \tiny{(246)} & 0.2632 \tiny{(76)} & 0.4187 \tiny{(246)} & 0.3421 \tiny{(76)} \\
JPN-Japan & 0.3371 \tiny{(175)} & 0.3750 \tiny{(56)} & \textbf{0.4011} \tiny{(177)} & 0.2453 \tiny{(53)} & 0.3503 \tiny{(177)} & 0.3962 \tiny{(53)}\\
LAT-Latvia & 0.3214 \tiny{(28)} & 0.5556 \tiny{(9)} & 0.2188 \tiny{(32)} & 0.2500 \tiny{(8)} & 0.4688 \tiny{(32)} & 0.5000 \tiny{(8)} \\
ITA-Italy & 0.3206 \tiny{(209)} & 0.4200 \tiny{(50)} & 0.2712 \tiny{(236)} & 0.2727 \tiny{(66)} & 0.4703 \tiny{(236)} & 0.3788 \tiny{(66)} \\
RCF-Rcf & 0.3091 \tiny{(110)} & 0.3333 \tiny{(27)} & 0.3017 \tiny{(116)} & 0.4074 \tiny{(27)} & 0.3707 \tiny{(116)} & 0.2222 \tiny{(27)} \\
USA-United States & 0.3077 \tiny{(234)} & 0.3676 \tiny{(68)} & 0.2989 \tiny{(271)} & 0.2840 \tiny{(81)} & 0.4502 \tiny{(271)} & 0.3704 \tiny{(81)} \\
CHN-China & 0.3068 \tiny{(176)} & 0.3774 \tiny{(53)} & 0.3333 \tiny{(192)} & 0.2653 \tiny{(49)} & 0.3750 \tiny{(192)} & 0.4286 \tiny{(49)}\\
NOR-Norway & 0.2915 \tiny{(247)} & 0.4750 \tiny{\textbf{(80)}} & 0.3377 \tiny{\textbf{(305)}} & 0.3182 \tiny{\textbf{(88)}} & 0.4557 \tiny{\textbf{(305)}} & 0.3750 \tiny{\textbf{(88)}} \\
AUS-Australia & 
0.2806 \tiny{(196)} & 0.4000 \tiny{(55)} & 0.2878 \tiny{(205)} & 0.3214 \tiny{(56)} & 0.4049 \tiny{(205)} & 0.4107 \tiny{(56)} \\
BLR-Belarus & 
0.2800 \tiny{(25)} & 0.2857 \tiny{(7)} & 0.2632 \tiny{(19)} & 0.4000 \tiny{(5)} & 0.3158 \tiny{(19)} & 0.2000 \tiny{(5)}\\
FIN-Finland & 
0.2745 \tiny{(153)} & 0.3404 \tiny{(47)} & 0.2614 \tiny{(153)} & 0.2250 \tiny{(40)} & 0.3987 \tiny{(153)} & 0.2250 \tiny{(40)}
\\
GER-Germany & 
0.2743 \tiny{(175)} & 0.2955 \tiny{(44)} & 0.3046 \tiny{(151)} & 0.4000 \tiny{(45)} & 0.4106 \tiny{(151)} & 0.3333 \tiny{(45)}
\\
ENG-England & 
0.2712 \tiny{(118)} & 0.3056 \tiny{(36)} & 0.3028 \tiny{(109)} & 0.2308 \tiny{(26)} & 0.2752 \tiny{(109)} & 0.2692 \tiny{(26)}\\
TUR-T\"urkiye & 
0.2703 \tiny{(148)} & 0.2381 \tiny{(42)} & 0.3333 \tiny{(123)} & 0.1795 \tiny{(39)} & 0.3333 \tiny{(123)} & 0.3846 \tiny{(39)}\\
HUN-Hungary & 
0.2667 \tiny{(120)} & 0.5000 \tiny{(40)} & 0.2441 \tiny{(127)} & 0.3235 \tiny{(34)} & 0.3543 \tiny{(127)} & 0.2059 \tiny{(34)} \\
KOR-Korea & 
0.2621 \tiny{(206)} & 0.3710 \tiny{(62)} & 0.2906 \tiny{(203)} & 0.4407 \tiny{(59)} & 0.3990 \tiny{(203)} & 0.3051 \tiny{(59)}\\
EST-Estonia & 
0.2538 \tiny{(197)} & 0.3898 \tiny{(59)} & 0.3048 \tiny{(187)} & 0.2963 \tiny{(54)} & 0.4439 \tiny{(187)} & 0.3704 \tiny{(54)}\\
CZE-Czech Republic & 
0.2500 \tiny{(224)} & 0.3051 \tiny{(59)} & 0.3227 \tiny{(220)} & 0.3833 \tiny{(60)} & 0.4182 \tiny{(220)} & 0.2500 \tiny{(60)}
\\
IRL-Ireland & 
0.2424 \tiny{(33)} & 0.5000 \tiny{(10)} & 0.1795 \tiny{(39)} & 0.1667 \tiny{(12)} & \textbf{0.5128} \tiny{(39)} & 0.3333 \tiny{(12)}\\
ISR-Israel & 
0.2273 \tiny{(22)} & \textbf{0.5714} \tiny{(7)} & 0.2941 \tiny{(17)} & 0.4000 \tiny{(5)} & 0.2353 \tiny{(17)} & 0.0000 \tiny{(5)}
\\
DEN-Denmark & 
0.2057 \tiny{(141)} & 0.1500 \tiny{(40)} & 0.3333 \tiny{(126)} & 0.2564 \tiny{(39)} & 0.4444 \tiny{(126)} & 0.4103 \tiny{(39)}
\\
ESP-Spain & 
0.2037 \tiny{(162)} & 0.2917 \tiny{(48)} & 0.3267 \tiny{(150)} & 0.3158 \tiny{(38)} & 0.3267 \tiny{(150)} & 0.3421 \tiny{(38)}\\
AUT-Austria & 
0.1852 \tiny{(54)} & 0.4737 \tiny{(19)} & 0.2381 \tiny{(63)} & \textbf{0.4667} \tiny{(15)} & 0.3810 \tiny{(63)} & 0.0667 \tiny{(15)}
\\
SVK-Slovakia & 
0.1818 \tiny{(22)} & 0.4000 \tiny{(10)} & 0.2903 \tiny{(31)} & 0.1429 \tiny{(7)} & 0.3226 \tiny{(31)} & 0.5714 \tiny{(7)}
\\
NED-Netherlands & 
0.1759 \tiny{(108)} & 0.1333 \tiny{(30)} & 0.3103 \tiny{(87)} & 0.2593 \tiny{(27)} & 0.3333 \tiny{(87)} & 0.3704 \tiny{(27)}\\
FRA-France & 
0.1698 \tiny{(53)} & 0.2000 \tiny{(15)} & 0.2083 \tiny{(48)} & 0.2000 \tiny{(10)} & 0.4375 \tiny{(48)} & \textbf{0.6000} \tiny{(10)}\\
POL-Poland & 0.1500 \tiny{(20)} & 0.2857 \tiny{(7)} & 0.2941 \tiny{(17)} & 0.1667 \tiny{(6)} & 0.2941 \tiny{(17)} & 0.1667 \tiny{(6)}\\
            \hline
\textbf{Weighted Average} 
&  \textbf{0.2820} \tiny{(119)} & \textbf{0.358} \tiny{(34)} 
& \textbf{0.3127} \tiny{(119)} & \textbf{0.3034} \tiny{(34)} 
& \textbf{0.4053} \tiny{(119)} & \textbf{0.3402} \tiny{(34)}\\
            \hline\\
 	\end{tabular}
 	\caption{For the top 30 countries sorted by Hammer efficiency, we present their average Hammer efficiency (HE), Force efficiency (FE) and Steal efficiency (SE) within Power Play (PP) and regular ends. In parentheses next to the efficiencies are the counts of ends the country possessed the hammer or not, which are the denominators for the efficiency proportions. Bold values are the highest within their column and the averages are presented in the final row weighted by game counts.
    }
 	\label{tab:hammer_efficiency}
 \end{table}  

Interestingly, few teams maintain above-average performance across all efficiency types, highlighting the complexity of mastering both offensive and defensive play, especially with the added strategic layer introduced by the Power Play. The relatively low sample size per team---at most one offensive and one defensive Power Play opportunity per game---limits opportunities for in-game learning and optimization. Looking closer at the Power Play, teams are expected to have a higher Hammer efficiency with the Power Play than without. For many teams, such as Switzerland, T\"urkiye, Denmark and the Netherlands, this is not the case. Perhaps these teams prefer to use the Power Play in a defensive manner to limit steals or simply take one point, but typically we expect higher values here. Similarly with the Steal efficiency, we expect a lower SE in the Power Play than otherwise, but for teams like Japan, Latvia, and China, their PP-SE is greater than SE. Perhaps these teams have a defensive edge in the Power Play that other teams are not prepared for.

Compared to traditional curling, the mixed doubles game reveals marked differences. \cite{Palmer2019} report Force Efficiencies between 45\%–65\% in traditional curling, whereas the highest in mixed doubles is just 40\% even though we expanded the definition to include blanks. This is likely due to the higher frequency of steals in the mixed doubles format. Similarly, Hammer Efficiencies in standard curling range from 30\%–50\%, while the weighted average in mixed doubles is approximately 28\%, indicating a scoring environment that is generally more evenly distributed among competitors.

These insights reinforce the conclusion that mixed doubles curling requires distinct strategic approaches and that modeling efforts must be tailored to this format, rather than relying on frameworks developed for traditional 4-player curling.



\section{Summary of Contributions}\label{Section:Summary}
In this work, we present the most comprehensive and structured dataset to date for mixed doubles curling, a sport that continues to evolve both competitively and analytically. We detail the processes involved in data collection, cleaning, and feature engineering, resulting in a dataset that spans nearly a decade, with over 60,000 shots, 1,100 games, and representation from 53 countries.

Our brief summary analysis highlights the unique dynamics of mixed doubles curling—differentiating it from traditional curling in key areas such as scoring distributions, shot success rates, and team strategies. We explore this through a multi-level analysis:
\begin{itemize}
    \item At the shot level, we discuss shot types and grading results, highlighting the relative success rates of different throws.
    \item At the end level, we analyze scoring patterns and the strategic impact of the Power Play, uncovering how it shifts scoring proportions and changes steal likelihoods across game contexts.
    \item At the game level, we aggregate performance metrics such as draw and take-out success rates to identify traits of top-performing teams.
    \item At the team level, we extend established standard curling metrics (e.g., Hammer, Force, and Steal Efficiencies) to the mixed doubles format and quantify their values in both standard and Power Play ends.
\end{itemize}

Our findings reveal that mixed doubles curling exhibits distinct structural and strategic differences compared to traditional curling. For example, the higher rate of steals, the near absence of blank ends, and the tactical use of the Power Play create novel decision-making challenges for teams. These elements make the findings from traditional curling analytics inappropriate for mixed doubles, reinforcing the need for tailored analytical tools.

We envision this dataset as a launching point for a wide range of future research. Potential directions include optimal Power Play timing and defense, role-based performance modeling (e.g., stone order assignments), in-game win probability estimation, and broader player/team evaluation frameworks. By making this data available and demonstrating its analytical potential, we aim to support the continued growth of mixed doubles curling through evidence-based strategies and performance insights.


\newpage
\bibliography{curling.bib}

\newpage
\section{Supplementary Material}\label{Section:SupMat}
\subsection{Text Scraping Corrections}\label{Sup:Textscraping}
\begin{figure}[H]
    \begin{center}
        \includegraphics[width=0.85\linewidth]{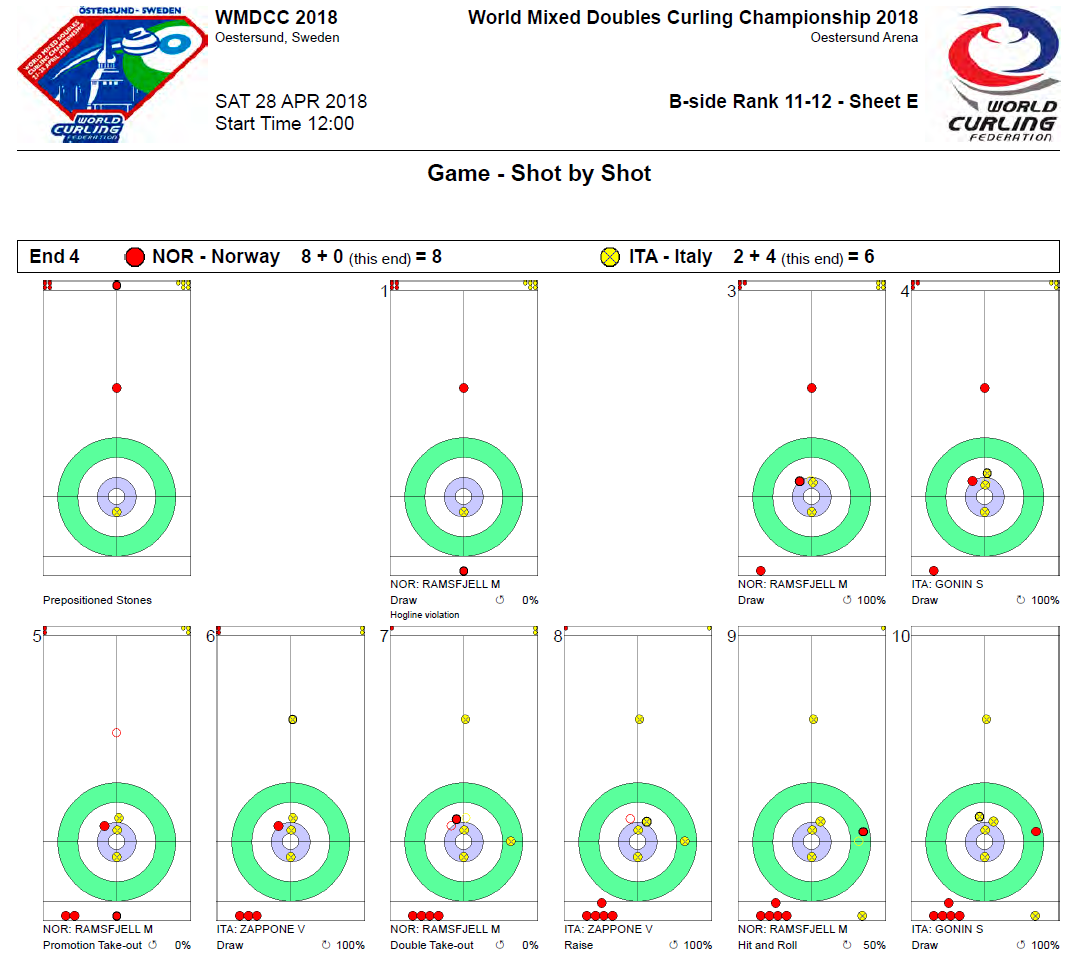}
        \caption{Example of a missing shot in the WMDCC 2018 tournament game betweek Norway and Italy. Data for the row in the shot-level dataset is nullified for impacted fields.}
        \label{fig:missing_shot}
    \end{center}
\end{figure}

Although most tournaments followed consistent formatting, several exceptions required manual intervention to ensure a complete and standardized dataset. Below we summarize the main categories of corrections applied.
\begin{enumerate}
    \item Tournament Metadata
    \begin{itemize}
        \item Some title pages were embedded as images and could not be parsed automatically through text scraping; tournament name, location, and dates were entered manually.
    \end{itemize}
    \item Roster, Player and Staff Information
    \begin{itemize}
        \item Name formatting: Inconsistent capitalization or abbreviation occasionally disrupted parsing (e.g., McFADZEAN with lowercase “c”).
        \item Non-unique identifiers: Cases where two players shared the same abbreviation (e.g., Maia and Magnus Ramsfjell both listed as RAMSFJELL M) required external verification via video sources.
        \item Abbreviated first names: Some entries (e.g., HAMILTON B for Rebecca “Becca” Hamilton) were harmonized to ensure consistent player tracking across tournaments.
        \item Missing attributes: Handedness varied across years; in such cases, we standardized fields and coded unavailable values as missing.
    \end{itemize}
    \item Game Results Pages
    \begin{itemize}
        \item Power Play inconsistencies: In some events (e.g., 2016 WMDCC), Power Plays were not identified or were mis-attributed. These were verified using shot by shot images and hammer possession.
        \item Shot success analysis: Early Results Books (e.g., 2016 WMDCC) lacked rotation-specific breakdowns. These omissions were flagged and coded as missing.
        \item Formatting issues: Long player names were occasionally split across lines, requiring manual reassembly.
    \end{itemize}
    \item Shot by shot Data
    \begin{itemize}
        \item Partial availability: In some years (e.g., 2018–2019 WMDCC), shot by shot results were only recorded for playoff rounds.
        \item Ambiguous shooters: Instances of mislabeled or blank shooter fields were corrected manually (e.g., “Power Play” erroneously listed as the shooter, or missing shooter names).
        \item Missing shots: Rare cases of absent shot records (e.g., a missing shot 2 in the 2018 WMDCC, NOR–ITA, End 4 shown in Figure~\ref{fig:missing_shot}) were filled with null entries to preserve sequencing.
        \item Rotation and rating formats: Differences in notation (e.g., In/Out vs. Clockwise/Counterclockwise) and rating scales (0–4 vs. 0–100\%) were standardized.
    \end{itemize}
\end{enumerate}

\subsection{Image Extraction Corrections}\label{Sup:Image_Issues}
\begin{figure}[H]
    \begin{center}
        \includegraphics[width=0.85\linewidth]{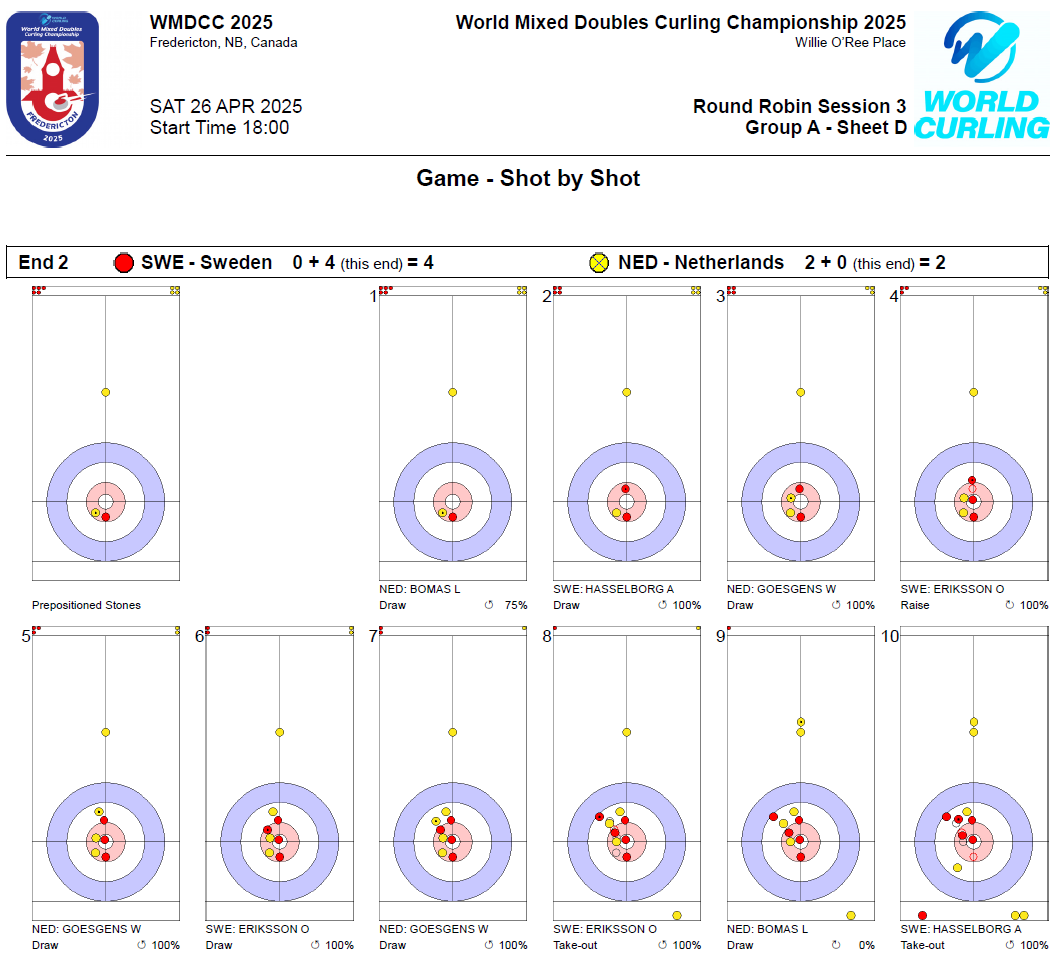}
        \caption{Example of a duplicate shot where the prepositioned stones images is a copy of the shot number 1. This is a page from the WMDCC 2025 tournament between Sweden and the Netherlands. To correct, we replace the incorrect image with an appropriate shot from another end before the processing step.}
        \label{fig:Dup_shot}
    \end{center}
\end{figure}

Table~\ref{tab:Preposition_Stone_Fixes} documents the tournaments, games, ends, and teams where manual corrections to shot by shot images were necessary. The most common issues were:
\begin{itemize}
    \item \textbf{Missing images} – prepositioned stone or regular shot images absent from the Results Booklets.
    \item \textbf{Duplicated images} – the prepositioned stones image duplicated from the first shot image.
\end{itemize}
Examples of the pages with missing or incorrect shots are show in Figures~\ref{fig:missing_shot} and \ref{fig:Dup_shot}.

\begin{table}
 	\centering
    \scalebox{0.9}{
 	\begin{tabular}{llllll}
 		\hline
 		Tournament & Game & End & Team 1 & Team 2 & Description \\
 		\hline
 		WMDCC 2017 & Round Robin Session 4 & 5 & Switzerland & Romania & Prepositioned image duplicate of shot 1\\
            WMDCC 2017 & Round Robin Session 8 & 5 & T\"urkiye & Czech Republic & Prepositioned image duplicate of shot 1\\
            WMDCC 2017 & Round Robin Session 9 & 8 & Israel & Australia & Prepositioned image duplicate of shot 1\\
            WMDCC 2017 & Round Robin Session 13 & 3 & Brazil & Hungary & Prepositioned image duplicate of shot 1\\
            WMDCC 2017 & Round Robin Session 17 & 5 & Spain & Switzerland & Prepositioned image missing\\
            WMDCC 2017 & Round Robin Session 21 & 5 & Finland & Brazil & Prepositioned image duplicate of shot 1\\
            WMDCC 2017 & Round Robin Session 23 & 6 & RCF & New Zealand & Prepositioned image duplicate of shot 1\\
            WMDCC 2017 & Round Robin Session 26 & 3 & Slovenia & Romania & Prepositioned image duplicate of shot 1\\
            WMDCC 2017 & Round Robin Session 26 & 6 & Norway & Ireland & Prepositioned image duplicate of shot 1\\
            WMDCC 2017 & Round Robin Session 27 & 3 & New Zealand & Switzerland & Prepositioned image missing\\
            WMDCC 2018 & B-side Rank 11-12 & 4 & Norway & Italy & Shot 2 missing\\
            OWG 2018 & Round Robin Session 1 & 5 & United States & OAR & Prepositioned image missing\\
            OWG 2018 & Semi-final & 8 & Canada & Norway & Prepositioned image missing\\
            WMDCC 2022 & Round Robin Session 13 & 3 & Italy & Switzerland & Prepositioned image missing\\
            WMDCC 2022 & Round Robin Session 18 & 2 & Germany & England & Prepositioned image missing\\
            WMDCC 2024 & Round Robin Session 14 & 2 & China & Canada & Prepositioned image missing\\
            WMDCC 2025 & Round Robin Session 2 & 8 & Australia & Switzerland & Prepositioned image missing\\
            WMDCC 2025 & Round Robin Session 3 & 2 & Sweden & Netherlands & Prepositioned image duplicate of shot 1\\
 		\hline\\
 	\end{tabular}
    }
 	\caption{List of manual shot by shot corrections required for the various tournaments excluding the WMDCC 2016.}
 	\label{tab:Preposition_Stone_Fixes}
 \end{table}
 
\subsection{Supplementary Material: Column Definitions}\label{Sup:columns}
The game- and shot-level dataset column descriptions are provided in Tables~\ref{Tab:Game_Columns} and \ref{Tab:Shot_Columns}. Options within \{\} are used to identify multiple columns of that type. For example T\{1,2\} identifies that there is one column called T1 and another called T2 representing the two teams playing in the game. 

\subsection{Supplementary Material: Scoring Distributions}\label{Sup:Scoring}
\begin{sidewaysfigure}
    \begin{center}
        \includegraphics[width=1\linewidth]{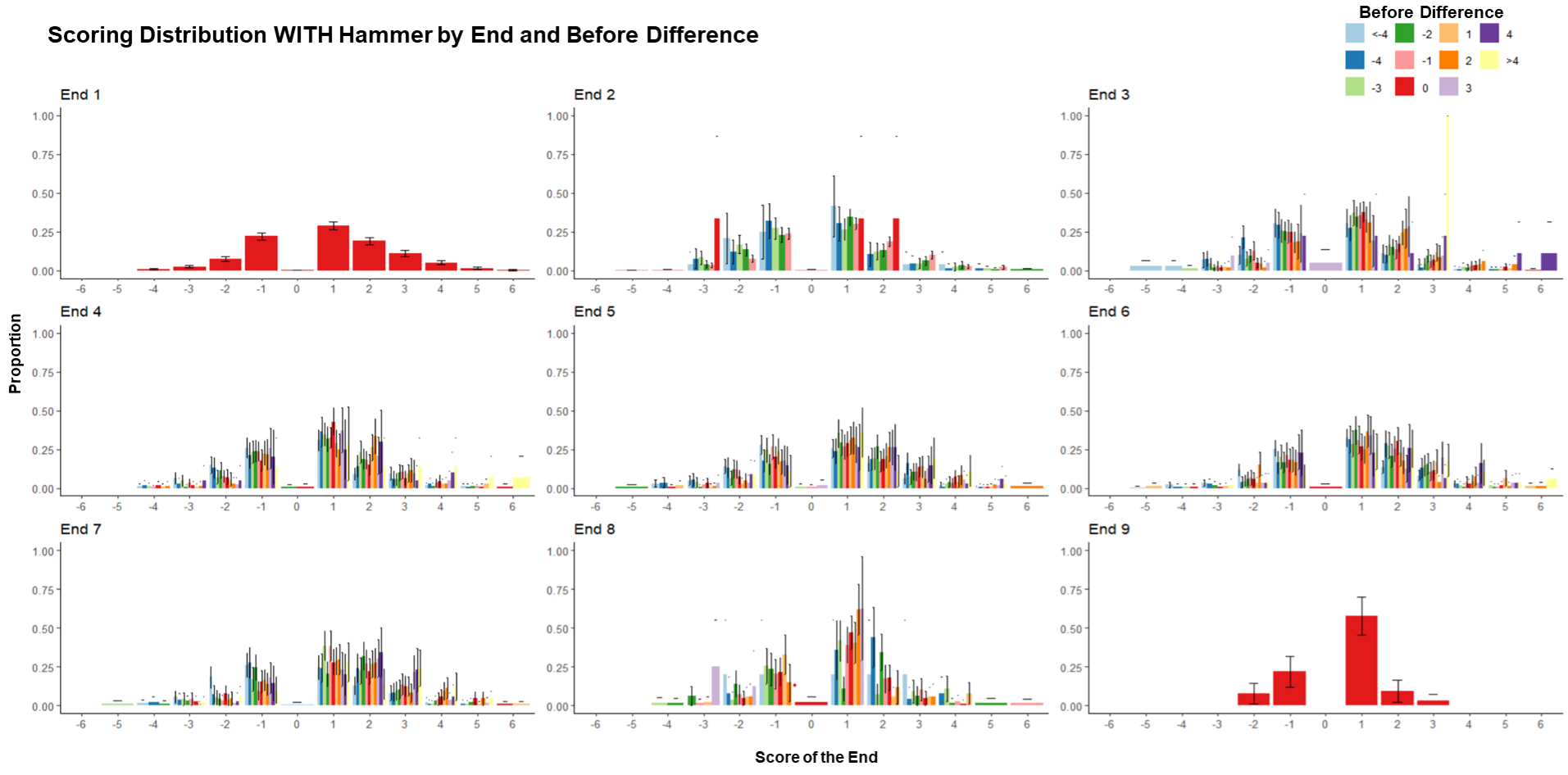}
        \caption{Scoring distribution by end and score differential prior to the end (``Before Difference") for the hammer team. }
        \label{fig:mx_scoring_dist_b4diff_end}
    \end{center}
\end{sidewaysfigure}

\begin{sidewaysfigure}
    \begin{center}
        \includegraphics[width=1\linewidth]{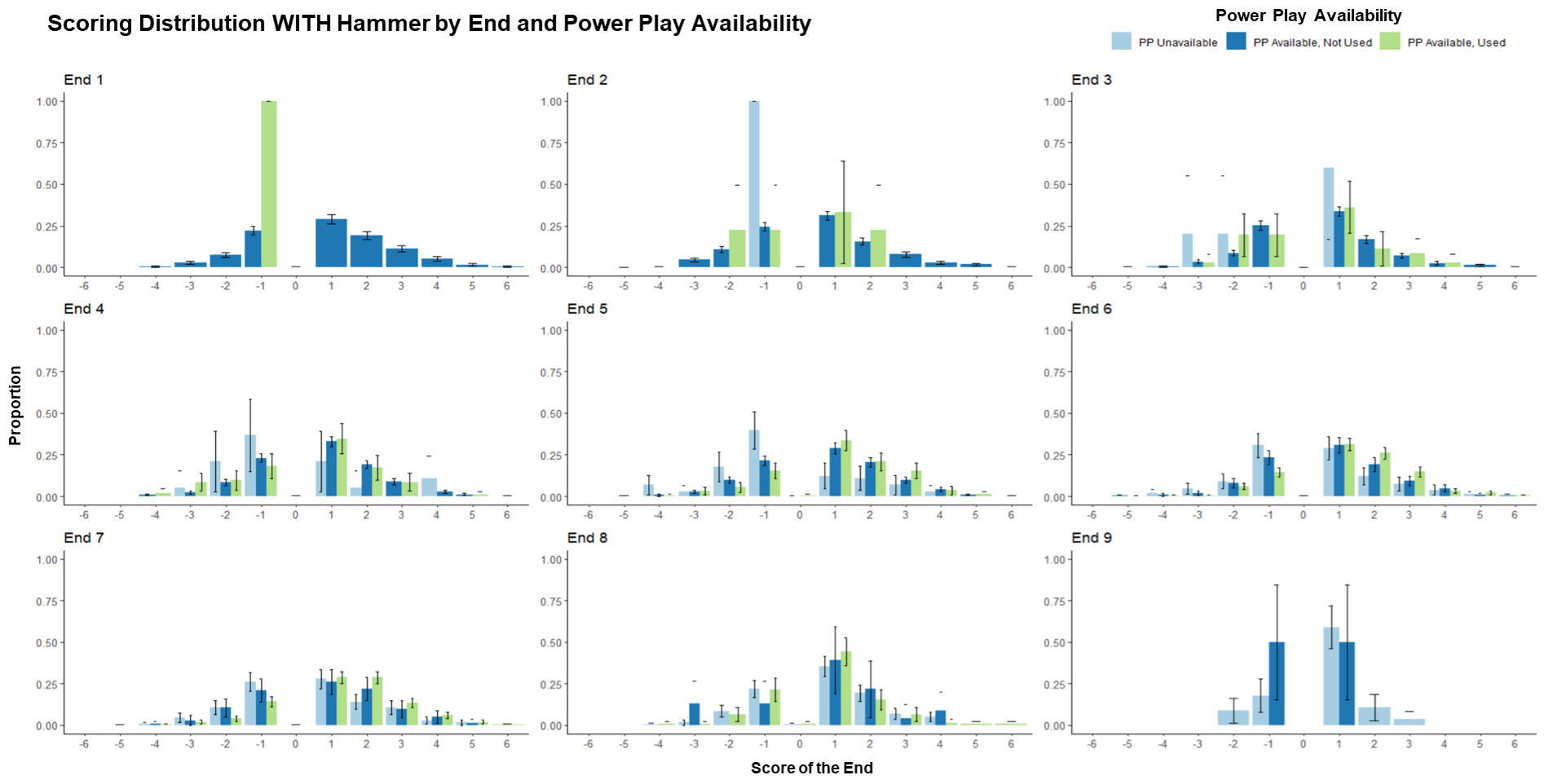}
        \caption{Scoring distribution by end and Power Play availability for the hammer team.}
        \label{fig:mx_scoring_dist_pp_end}
    \end{center}
\end{sidewaysfigure}

In addition to the figures provided in Section~\ref{EndAnalysis}, we further split the scoring distributions for the hammer team by score differential and end (Figure~\ref{fig:mx_scoring_dist_b4diff_end}) and by score differential and Power Play usage (Figure~\ref{fig:mx_scoring_dist_pp_end}). Sample size has an effect on the results, which is visible when separating the data by Power Play usage. For example, only one team has used the Power Play in end 1 and, thus, was the only team to have the advantage unavailable in the second end.

\scalebox{0.7}{
\rotatebox{90}{   
 	\begin{tabular}{ll}
            Game by game dataset details\\
 		\hline
 		Column & Description \\
 		\hline
 	 ID &  Tournament identifier, a combination of the abbreviation of the tournament and the year.\\
          Title & Name of the game, typically round robin or playoff label.\\
          Game & Increment number for the game, gold medal game is game 1 and the largest number is first round robin game of the tournament.\\
          Sheet & Sheet label (when available).\\
          T\{1,2\} & Team names. \\
          LineScoreT\{1,2\} & End by end breakdown of what the team scored.\\
          FinalT\{1,2\} &  Score at the conclusion of the game.\\
          LSFE & Indicator for Last Stone First End winner.\\
          T\{1,2\}\_Colour & Stone colour of each team. \\
          PP\_E\{1,2\} & Identifiers for the ends that had a Power Play.\\
          Left\_PP\_E\{1,2\} & Indicator for whether the Power Play was on the left side of the sheet.\\
          T\{1,2\}\{F,M\}Name & Name of team member by gender.\\
          T\{1,2\}\{F,M\}\{Draws,TO,Total\}\{Clock,Counter\}Num & Number of shots per team member per shot type or total for the tournament thus far separated by stone rotation.\\
          T\{1,2\}\{F,M\}\{Draws,TO,Total\}\{Clock,Counter\}Percent & Outcome percentages of shots per team member per shot type or total for the tournament thus far separated by stone rotation.\\
          T\{1,2\}\{F,M\}\{Draws,TO,Total\}GameNum & Number of total shots per team member per shot type or total for the tournament thus far.\\
          T\{1,2\}\{F,M\}\{Draws,TO,Total\}GamePercent & Outcome percentages of total shots per team member per shot type or total for the tournament thus far.\\
          T\{1,2\}\{Draws,TO,Total\}\{Clock,Counter\}Num & Number of shots per team per shot type or total for the  tournament thus far separated by stone rotation.\\
          T\{1,2\}\{Draws,TO,Total\}\{Clock,Counter\}Percent & Outcome percentages of shots per team per shot type or total for the tournament thus far separated by stone rotation.\\
          T\{1,2\}\{Draws,TO,Total\}GameNum & Number of total shots per team per shot type or total for the tournament thus far.\\
          T\{1,2\}\{Draws,TO,Total\}GamePercent & Outcome percentages of total shots per team per shot type or total for the tournament thus far.\\
          T\{1,2\}\{F,M,Total\}Game & Number of shots per team member for the game.\\
          T\{1,2\}\{F,M,Total\}All &  Outcome percentages of shots per team member for the tournament thus far.\\
          T\{1,2\}Mins & Minutes remaining on the shot clock at the end of the game by team.\\
          T\{1,2\}Secs & Seconds remaining on the shot clock at the end of the game by team.\\
          EE-T\{1,2\}Mins & Extra ends minutes remaining on the shot clock at the end of the game by team.\\
          EE-T\{1,2\}Secs & Extra ends seconds remaining on the shot clock at the end of the game by team.\\
          T\{1,2\}Time-out\{1,2\}\_End & Time-out end that the team called each of their two time-outs.\\
          T\{1,2\}Time-out\{1,2\}\_Shot & Time-out shot number that the team called each of their two time-outs.\\
          T\{1,2\}\_\{F,M\}\_LSFE\_Turn & Last stone draw rotation per team member.\\
          T\{1,2\}\_\{F,M\}\_LSFE\_Distance & Last stone draw distance to the pin per team member.\\
          T\{1,2\}\_LSFE\_TotalDistance & Team's cumulative last stone draw distance to the pin.\\
 		\hline\\
 	\end{tabular}
 	\label{Tab:Game_Columns}
 }}
  
\scalebox{0.6}{
\rotatebox{90}{    
 	\begin{tabular}{ll}
            Shot by Shot dataset details\\
 		\hline
 		Column & Description \\
 		\hline
 	 ID &  Tournament identifier, a combination of the abbreviation of the tournament and the year.\\
          Title & Name of the game, typically round robin or playoff game.\\
          Game & Increment number for the game, gold medal game is game 1 and the largest number is first round robin game of the tournament.\\
          Sheet & Sheet label (when available).\\
          T\{1,2\} & Team names.\\
          T\{1,2\}\_Colour & Stone colour of each team.\\
          T\{1,2\}\{F,M\}Name & Names of the female and male members of each team.\\
          LSFE & Indicator for last stone first end winner.\\
          End & Number of the end.\\
          Hammer\_End & Name of the team with hammer possession in the end.\\
          Shot & Number of the shot, 0 indicates prepositioned stones.\\
          Rock\_Country & Name of the team throwing the stone.\\
          Rank & Final standing ranking, 1 indicates the tournament winner, -1 indicates the team was disqualified.\\
          Gender & Gender of the player throwing the shot.\\
          Thrower & Name of the player throwing the shot.\\
          Handedness & Handedness of the thrower.\\
          Throw & Type of throw (e.g. Draw, Take-out, etc.).\\
          Turn & Rotation of the throw.\\
          Percent & Score of the shot (0\%, 25\%, 50\%, 75\%, or 100\%).\\
          \{R,Y\}\{1,2,3,4,5,6\}\{X,Y\} & Shot identifier, combination of the rocks colour, number and Y or X coordinate. The number is not necessarily the order it was shot but rather the order in which the algorithm identified it.\\
          ShooterCol & Colour of the shooter stone. The shooter is identified by a bold black outline or a black dot in its center. \\
          Shooter\{X,Y\}& Coordinates of the shooter stone.\\
          \{Left,Right\}\_PP& Indicator for whether the Power Play has on the right or left side of the sheet.\\
          PowerPlay & Indicator for if the end is a Power Play.\\
          PP\_Team & Identifier for the team that called the Power Play and hence had prepositioned stone placement selection in that end.\\
          T\{1,2\}\_has\_PP\_available & Indicator for whether or not the team has a Power Play available during that end. The indicator is 1 during the Power Play end and also even when the team doesn't have the hammer and are unable to use it in the given end.\\
          T\{1,2\}Before & The score for the team prior to the end.\\
          T\{1,2\}During & The amount of points the team scored as a result of the end. X if the game was conceded during the end.\\
          T\{1,2\}After & T\{1,2\}Before + T\{1,2\}During. X if the game was conceded during the end.\\
          FinalT\{1,2\} & Score at the conclusion of the game. \\
          \{Red,Yellow\}\_Score & Number of each colours stones in the house and closer to the pin than the opposite colour, thus the score at the conclusion of the shot.\\
          \{Red,Yellow\}\_House & Number of each colours stones in the house.\\
          \{Red,Yellow\}\_Button & Number of each colours stones on the button. \\
          \{Red,Yellow\}\_\{4,8,12\}FT & Number of each colours stones in the ring of that size.\\
          \{Red,Yellow\}\_W\{1,...,10\} & Number of each colours stones in the playing zone split by weight class, see Figure~\ref{fig:weights}.\\
          \{Red,Yellow\}\_G\{L,C,R\} &  Number of each colours stones in the playing zone split by guard side, see Figure~\ref{fig:weights}.\\
 		\hline\\
 	\end{tabular}
    \label{Tab:Shot_Columns}
     }}

\end{document}